\documentclass[onefignum,onetabnum]{siamonline250211}

\usepackage[english]{babel}

\usepackage{amsmath}
\usepackage{amssymb}
\usepackage{amsfonts}

\usepackage{graphicx}
\usepackage{xcolor}

\usepackage{soul}

\usepackage{subcaption}
\usepackage{array}
\usepackage{multirow}

\usepackage{algorithm}
\usepackage{algpseudocode}
\usepackage{url}
\usepackage{tikz}
\usetikzlibrary{arrows.meta,positioning}

\ifpdf
  \DeclareGraphicsExtensions{.eps,.pdf,.png,.jpg}
\else
  \DeclareGraphicsExtensions{.eps}
\fi

\usepackage{enumitem}
\setlist[enumerate]{leftmargin=.5in}
\setlist[itemize]{leftmargin=.5in}

\tikzstyle{process} = [
  rectangle,
  draw=black,
  minimum width=3.5cm,
  minimum height=0.9cm,
  text centered,
  font=\small,
  fill=none
]

\tikzstyle{arrow} = [
  thick,
  ->,
  >=Stealth
]

\headers{B-GRASP: A Bayesian Framework for Inferring Graph Weights}
{C. Schenk and B. M. Afkham}

\title{
B-GRASP: A Bayesian Framework for Inferring Graph Weights
from SPDE-Inspired Dynamics
\thanks{
Submitted to the editors September 30, 2026.
\funding{
B. M. Afkham was partially supported by the
Flagship of Advanced Mathematics for Sensing, Imaging and Modelling
(Research Council of Finland grant no.~359186)
and by the Research Council of Finland under grant no.~371523. C. Schenk acknowledges financial support from the Spanish Ministry of Science and Innovation through a Ram\'on y Cajal program (Grant No. RYC2024-048744-I), funded by  MICIU/AEI/10.13039/501100011033 and FSE+.
}
}
}

\author{
Christina Schenk\textsuperscript{**}
\thanks{
IMDEA Materials Institute,
Eric Kandel 2, Getafe, 28906, Spain
(\email{christina.schenk@imdea.org}).
ORCID: 0000-0002-7817-6757.
}
\and
Babak Maboudi Afkham\textsuperscript{**}
\thanks{
University of Oulu, Oulu, Finland
(\email{babak.maboudi@oulu.fi}).
ORCID: 0000-0003-3203-8874.
}
}

\ifpdf
\hypersetup{
  pdftitle={An Example Article},
  pdfauthor={D. Doe, P. T. Frank, and J. E. Smith}
}
\fi

\begin{document}

\maketitle
\begingroup
\renewcommand{\thefootnote}{**}
\footnotetext{The authors contributed equally to this work.}
\endgroup
\begin{abstract}
We present B-GRASP (Bayesian GRAph inference with SPDE priors), a Bayesian framework for inferring uncertain edge weights in stochastic dynamical systems on graphs from noisy observations of nodal states. The unknown edge weights parameterize the graph differential operator and therefore directly govern the evolution of the graph process. Motivated by connections between differential operators in PDEs and SPDEs and their graph counterparts, we construct stochastic graph models incorporating diffusion, reaction dynamics, and stochastic forcing.
The resulting hierarchical formulation jointly represents uncertainty in the graph structure and stochastic forcing. Latent graph variables determine positive edge weights and the corresponding graph Laplacian, while latent Brownian variables represent the stochastic forcing. Conditional on these variables, the graph dynamics define a deterministic forward map from which the likelihood and posterior distribution are constructed. We characterize the posterior using maximum a posteriori estimation and the No-U-Turn Sampler, enabling both point estimation and uncertainty quantification.
We demonstrate the framework on a one-dimensional inverse heat-conduction problem, stationary and nonstationary graph reaction–diffusion systems with nonlinear dynamics, and state-level COVID-19 data in the United States. The numerical results show that posterior uncertainty provides information not captured by point estimates, particularly for weakly identifiable or highly conductive edges, and enables uncertainty in both graph connectivity and stochastic forcing to be quantified within a Bayesian framework.
\end{abstract}



\section{Introduction}
Graphs are a fundamental modeling tool across physics, engineering, and biology, capturing interactions within complex systems. As these systems grow, identifying their underlying connectivity becomes increasingly challenging, particularly when graph edge weights must be inferred from incomplete or noisy observations. Uncertainty in the inferred weights can significantly affect downstream analyses such as clustering, prediction, and control, motivating Bayesian approaches that incorporate prior information and quantify parameter uncertainty \cite{perri_bayesian_2023}.

Gaussian processes (GPs) provide a flexible framework for incorporating structured prior knowledge and uncertainty \cite{Rasmussen2006Gaussian}. Designing GP priors for heterogeneous graph domains, however, remains challenging because graph signals may vary over multiple spatial or spectral scales and graph domains may exhibit diverse connectivity, geometric, and topological properties. Covariance structures must therefore adapt to both graph topology and the observed process \cite{pmlr-v206-borovitskiy23arn,opolka_adaptive_2022,dunson2022graph}. Recent approaches include geometric analogues of heat and Matérn kernels on graphs, meshes, and manifolds \cite{mostowsky_geometrickernels_2025}, adaptive spectral graph wavelet kernels \cite{opolka_adaptive_2022}, data-driven spectral kernel learning \cite{zhi_gaussian_2023}, and Gaussian processes on cellular complexes that capture higher-order interactions \cite{alain_gaussian_2024}. Nevertheless, variations in graph size, connectivity, node attributes, and uncertainty in graph weights or structure continue to complicate the construction of broadly applicable priors \cite{Borovitskiy2021,goldberg1997regression,blanco2021evolving,rasmussen2004gaussian}.

A complementary route to constructing GP priors arises from stochastic partial differential equations (SPDEs). Beginning with Whittle \cite{whittle1963} and subsequently developed into practical computational frameworks \cite{Lindgren2011,Lindgren2015,solin_stochastic_2016}, this approach relates GP covariance structures to differential operators driven by stochastic forcing. Extensions include non-separable advection–diffusion models and non-stationary formulations with spatially varying coefficients \cite{berild_non-stationary_2024,clarotto_spde_2024}. For suitable covariance operators, GP realizations can therefore be characterized through corresponding SPDEs, providing a physical interpretation of covariance parameters while enabling the use of numerical methods for differential equations. SPDE-based methodologies have consequently found applications in animal movement prediction \cite{hooten2017animal}, geographic information systems \cite{burrough2015principles}, and geostatistics \cite{moraga2017geostatistical}.

Transferring this operator-based perspective to graphs is non-trivial because differential operators act on continuous domains whereas graphs are discrete. Graph operators such as the graph Laplacian nevertheless provide discrete analogues of differential operators. Connections between discrete and continuum diffusion have been studied through mathematical well-posedness and numerical discretization \cite{SchenkPortilloRomero2023}, while spectral graph theory uses the graph Laplacian to construct embeddings and kernels that capture graph topology \cite{Belkin2003,Yan2007}. This perspective has led to diffusion- and random-walk-based graph kernels \cite{Smola2003}, connections between graph discretizations and classical Matérn kernels \cite{sanzalonso2021spdeapproachmaternfields}, and Whittle–Matérn SPDE constructions for fields defined continuously on metric-graph edges \cite{bolin_gaussian_2024}.

SPDE-based kernel constructions have also been extended directly to graph-structured spatio-temporal processes \cite{Nikitin2022}. Nikitin et al.\ construct non-separable kernels that encode interactions between graph structure and temporal dynamics through stochastic analogues of physical processes such as diffusion and transport. Harlim et al.\ \cite{Harlim_2022} use graph-based approximations of differential operators for inverse problems on manifolds with geometric constraints, while SPDE-inspired constructions have also been investigated for uncertainty estimation on graphs through connections between message passing and stochastic spatio-temporal dynamics \cite{xumarkovich_uncertainty_nodate}. These approaches combine graph structure, stochastic dynamics, and probabilistic uncertainty representations, but primarily quantify uncertainty in signals or quantities defined on a prescribed graph rather than infer the graph weights themselves.

Inference of the graph weights from incomplete or noisy observations is an ill-posed inverse problem because the observations may not uniquely or stably determine the graph parameters. Bayesian inference addresses this ill-posedness by combining prior information with observations while explicitly quantifying parameter uncertainty \cite{stuart2010inverse,schenk2026ac}. Related statistical inference has been developed for Gaussian Whittle–Matérn fields on prescribed metric graphs using their Markov structure \cite{bolin_statistical_2026}. In contrast, we treat the edge weights themselves as uncertain parameters that directly determine the graph differential operator governing the dynamics.

Motivated by this distinction, we develop B-GRASP (Bayesian GRAph inference with SPDE priors), a Bayesian framework for stochastic dynamical systems on weighted graphs in which the edge weights are uncertain model parameters. Latent graph variables determine positive edge weights and hence the graph differential operator governing the evolution of the nodal states. The construction exploits the correspondence between PDE/SPDE differential operators and their graph analogues, allowing diffusion, reaction dynamics, and stochastic forcing to retain interpretations closely related to their continuous-domain counterparts.

B-GRASP uses a hierarchical formulation that jointly represents uncertainty in graph structure and stochastic forcing. Latent Brownian variables determine realizations of stochastic forcing and, conditional on these and the graph variables, graph dynamics define a deterministic forward computation from which the observation likelihood is constructed. The framework accommodates stationary and non-stationary, linear and nonlinear reaction--diffusion dynamics and is modular in the observation operator and measurement model.

The resulting nonlinear, non-Gaussian posterior is characterized using MAP estimation and gradient-based MCMC \cite{gilks1996markov,robert2013monte,neal2011mcmc}, specifically the No-U-Turn Sampler (NUTS) \cite{hoffman2014no}, enabling complementary point estimation and posterior uncertainty quantification.

We investigate B-GRASP through examples of increasing complexity, from a one-dimen-sional inverse heat-conduction problem connected to classical finite-difference discretization to stochastic (non)-linear graph reaction–diffusion systems and a real-world application to state-level COVID-19 data in the United States. Together, these examples show that classical coefficient-estimation problems and stochastic nonlinear graph dynamics can be treated within a Bayesian framework. Importantly, the results demonstrate that point estimates may provide an incomplete characterization of uncertain graph connectivity: for highly conductive or weakly identifiable edges, MAP estimates and posterior means can underestimate the true weights, while the posterior distribution reveals a substantially broader range of values compatible with the observations. This distinction highlights the role of uncertainty quantification in identifying graph parameters that are only weakly constrained by the observed dynamics.

The remainder of the paper is organized as follows. Section 2 introduces the Bayesian inverse graph problem together with its SPDE-based formulation. Stochastic processes on weighted graphs and their temporal discretization are developed in Section 3. Section 4 then presents the complete B-GRASP framework and posterior inference, followed by the numerical results in Section 5. Finally, Section 6 summarizes the main findings.
\section{Inverse graph problems}\label{sec:invgraphprob}
In this section, we introduce graphs with time-dependent vertex values and varying weighted edges. We then formulate the inverse graph problems for inferring weights, introduce stochastic partial differential equations, which will serve as priors in the thereafter presented Bayesian formulation.

We denote a graph by $\mathcal G = (\mathcal V,\mathcal E)$, where $\mathcal V = \{ v_1\dots, v_{N_{\mathcal V}} \}$ are a set of vertices, and $\mathcal E \subset \mathcal V\times \mathcal V$ a set of $N_{\mathcal E}$ edges.
Each edge $(v_i,v_j)\in \mathcal E$, for some $1 \leq i,j \leq N_{\mathcal V}$, is associated with a weight $w_{ij}\in \mathbb R_{\geq 0}$, and we collect them in a vector $\boldsymbol w$. In this work, we assume that the weights are symmetric, i.e., $w_{ij} = w_{ji}$, for any $i$ and $j$ with $(v_i,v_j) \in \mathcal E$. We define the adjacency matrix $A \in \mathbb R_{\geq 0}^{N_{\mathcal V}\times N_{\mathcal V}}$, or $A(\boldsymbol{w})$ to highlight its dependence on $\boldsymbol{w}$, to be the symmetric matrix that holds $\boldsymbol{w}$.

To model time-dependent values at the vertices, we associate with each vertex $v_i$ a function $f_i : [0, T] \to \mathbb{R}$, where $T < \infty$ denotes the time horizon of interest. We collect the vertex values into the vector-valued function
\[
    \boldsymbol{f}(t;\boldsymbol{w}) : [0,T] \to \mathbb{R}^{N_{\mathcal V}}.
\]
The dependence on $\boldsymbol{w}$ emphasizes that the dynamics of the vertex values are governed by the weighted graph; the precise form of this dependence will be introduced below.

The inverse graph problem considered here consists of inferring the graph weights
$\boldsymbol{w}$ from finitely many noisy observations of $\boldsymbol{f}$. In
particular, we consider
\begin{equation}
    \label{eq:graph-inverse}
    \boldsymbol{y}_i
    =
    \boldsymbol{f}(i\Delta t;\boldsymbol{w})
    +
    \boldsymbol{\varepsilon}_i,
    \qquad i = 1,\dots,N_T,
\end{equation}
where $\Delta t$ denotes the time interval between consecutive measurements and
$\boldsymbol{\varepsilon}_i$ represents the additive Gaussian measurement noise and $N_T$ denotes the number of observation times.
The observations $\boldsymbol{y}_i \in \mathbb{R}^{N_{\mathcal V}}$ are collected
into $\boldsymbol{y} \in \mathbb{R}^{N_T \times N_{\mathcal V}}$.

The direct observation model in \eqref{eq:graph-inverse} is chosen to isolate and highlight the role of graph-based priors in the inverse problem. The framework developed below can also accommodate non-trivial observation operators and more general noise models. To specify how the graph weights $\boldsymbol{w}$ govern the evolution of $\boldsymbol{f}$, we next introduce the SPDE framework from which our graph
dynamics will be constructed.

\subsection{Bayesian inverse problems}
\label{sec:BIP}
We now reformulate the inverse graph problem in \eqref{eq:graph-inverse} within a Bayesian framework. Let $W$ be an $\mathbb{R}_{\geq 0}^{N_{\mathcal E}}$-valued random variable representing the unknown graph weights, and let $\{F(t;W)\mid t\in[0,T]\}$ be an $\mathbb{R}^{N_{\mathcal V}}$-valued stochastic process describing the corresponding evolution of the vertex values. Furthermore, let $E$ and $Y$ denote random variables representing the measurement noise and observations, respectively.

Evaluating the stochastic process at the measurement times $t_i=i\Delta t$, $i=1,\dots,N_T$, and collecting the resulting values, we write
\begin{equation}
    \label{eq:graph-inverse-statistical}
    \begin{aligned}
        Y = \mathcal{F}(W) + E, \text{ where,} \quad
        \mathcal{F}(W)
        =
        \left(
            F(\Delta t;W),
            \ldots,
            F(N_T\Delta t;W)
        \right).
    \end{aligned}
\end{equation}
collects the vertex values at the measurement times. Thus, $Y,E,\mathcal{F}(W)\in \mathbb{R}^{N_T\times N_{\mathcal V}}$. Equation \eqref{eq:graph-inverse-statistical} is the statistical counterpart of the inverse problem in \eqref{eq:graph-inverse}, in which the unknown graph weights, measurement noise, and observations are represented as random
variables.

Given an observed realization $\boldsymbol{y}$ of $Y$, the solution of the Bayesian inverse problem is the conditional random variable $W\mid Y=\boldsymbol{y}$, whose distribution is referred to as the \emph{posterior distribution}. Suppose that the distributions of $W$ and $E$ admit probability density functions $\pi_W(\boldsymbol{w})$ and $\pi_E(\boldsymbol{e})$, respectively. The density $\pi_W(\boldsymbol{w})$, referred to as the \emph{prior density}, encodes information about the graph weights before observing the data, while $\pi_E(\boldsymbol{e})$ specifies the statistical model for the measurement
noise.

Bayes' rule relates the posterior density to the likelihood and prior according to
\begin{equation}
    \label{eq:posterior}
    \pi_{(W\mid Y=\boldsymbol{y})}(\boldsymbol{w})
    =
    \frac{
        \pi_{(Y\mid W=\boldsymbol{w})}(\boldsymbol{y})
        \pi_W(\boldsymbol{w})
    }{
        \pi_Y(\boldsymbol{y})
    }
    \propto
    \pi_{(Y\mid W=\boldsymbol{w})}(\boldsymbol{y})
    \pi_W(\boldsymbol{w}).
\end{equation}
Here, $\pi_{(Y\mid W=w)}(\boldsymbol{y})$ denotes the likelihood, which quantifies the compatibility of the observed data with a given set of graph weights, while $\pi_Y(\boldsymbol{y})$ is the marginal density of the observations. Since $\pi_Y(\boldsymbol{y})$ is independent of $\boldsymbol{w}$, it acts as a normalization constant when the posterior density is viewed as a function of the unknown graph weights.

Consequently, characterizing the posterior requires specifying both the likelihood and the prior. While the likelihood follows from the observation and noise models, the construction of an appropriate prior for the graph weights requires additional modeling assumptions. In the following, we introduce SPDE-based constructions that will provide the foundation for the graph priors considered in this work.

\subsection{Stochastic partial differential equations}\label{sec:spdes}
SPDEs provide a flexible mechanism for constructing random fields with prescribed spatial and temporal dependence and, consequently, structured priors for inverse problems. In particular, randomness can be introduced into the governing differential equations through stochastic forcing or random coefficients. We first consider a time-dependent SPDE of the form
\begin{equation}
    \label{eq:PDE}
    \frac{\partial f}{\partial t}(t,x)
    =
    \mathcal{L}f(t,x) + \xi(t,x),
\end{equation}
where $f(t,x)$ is a random field, $\mathcal{L}$ is a spatial differential operator describing, e.g., diffusion, advection, or reaction, and $\xi(t,x)$ denotes stochastic forcing.

Time-independent random fields can analogously be defined through
\begin{equation}
    \label{eq:PDE-stationary}
    \mathcal{L}f(x) = \xi(x),
\end{equation}
where $\xi(x)$ denotes a spatial stochastic forcing term. In this setting, the differential operator $\mathcal{L}$ together with the distribution of $\xi$ determines the spatial dependence structure of the resulting random field. This construction is particularly useful for defining Gaussian random fields and will later provide the basis for constructing Gaussian priors on graphs.

Beyond diffusion, PDE and SPDE models may include reaction terms that describe local changes in the state of the system. In contrast to the diffusion operator, which describes interactions across the spatial domain, a reaction term acts locally on the value of the field. Extending \eqref{eq:PDE}, we consider SPDEs of the general form
\begin{equation}
    \label{eq:reaction-spde}
    \frac{\partial f}{\partial t}(t,x)
    =
    \mathcal{L}f(t,x)
    +
    \beta \mathcal{R}\bigl(f(t,x)\bigr)
    +
    \xi(t,x),
\end{equation}
where $\mathcal{L}$ describes spatial interactions, $\mathcal{R}$ denotes the reaction term, $\beta\in\mathbb{R}_{>0}$ represents the weight of the reaction term, and $\xi$ represents stochastic forcing. Depending on the underlying application, $\mathcal{R}$ may be linear or nonlinear and can describe, for example, growth, decay, saturation, or other state-dependent local dynamics. In this work, we focus on employing a nonlinear reaction term, while a corresponding linearized model is considered in the supplementary material (SM).

\subsubsection{The Whittle--Matérn differential operator}
\label{sec:diffterm}

Many PDE models describe spatial interactions through the Laplace operator $\Delta$, which in $\mathbb{R}^d$ is defined by
\begin{equation}
    \Delta
    =
    \sum_{i=1}^d
    \frac{\partial^2}{\partial x_i^2}.
\end{equation}
In diffusion-type models, the Laplacian describes local spatial interactions whose collective effect gives rise to diffusion over the spatial domain. The same operator also plays an important role in the construction of Gaussian random fields through SPDEs.

A prominent example is the Matérn family of Gaussian random fields. The Matérn covariance function in $\mathbb{R}^d$ can be written as
\begin{equation}
    k_{\mathrm{Mat\acute ern}}(x,x')
    =
    \sigma^2
    \frac{2^{1-\mu}}{\Gamma(\mu)}
    \left(
        \kappa \|x-x'\|
    \right)^\mu
    K_\mu
    \left(
        \kappa \|x-x'\|
    \right),
\end{equation}
where $\sigma^2>0$ determines the marginal variance, $\mu>0$ controls field smoothness, $\kappa>0$ is a spatial scale (inverse-range) parameter, and $K_\mu$ denotes the modified Bessel function of the second kind.

An important result due to Whittle is that a Gaussian random field with Matérn covariance can equivalently be characterized through an SPDE \cite{lindgren2011explicit}. In $\mathbb{R}^d$, this representation is
\begin{equation}
    \label{eq:whittle-matern-spde}
    \tau
    \left(
        \kappa^2-\Delta
    \right)^{\alpha/2}
    f(x)
    =
    \mathcal{W}(x),
\end{equation}
where $\mathcal{W}$ denotes Gaussian spatial white noise, $\tau>0$ is a scaling parameter, and $\alpha = \mu + \frac{d}{2}$. The parameters $\alpha$ and $\kappa$ determine the regularity and spatial dependence of the resulting random field, while $\tau$ controls its marginal variance.

The representation in \eqref{eq:whittle-matern-spde} is particularly useful because it characterizes the dependence structure of the random field through the differential operator
\begin{equation}
    \label{eq:whittle-matern-operator}
    \mathcal{L}_{\mathrm{WM}}
    =
    \left(
        \kappa^2-\Delta
    \right)^{\alpha/2},
\end{equation}
rather than directly through a covariance kernel. Formally,
\eqref{eq:whittle-matern-spde} may be written as
\begin{equation}
    f
    =
    \tau^{-1}
    \mathcal{L}_{\mathrm{WM}}^{-1}
    \mathcal{W},
\end{equation}
illustrating how spatial white noise is transformed by the inverse Whittle--Matérn operator into a spatially correlated Gaussian random field.

This operator-based characterization is central to the developments in this work. In particular, it provides a natural route for transferring SPDE-based Gaussian random field constructions from continuous spatial domains to finite-dimensional graph structures.
\section{Stochastic processes on graphs}
\label{sec:graph-processes}
We now transfer the SPDE constructions introduced above from continuous spatial domains to finite-dimensional weighted graphs. In the continuous setting, the state of the system is represented by a random field $f(t,x)$, whose spatial interactions are governed by a differential operator $\mathcal{L}$. On a graph with $N_{\mathcal V}$ vertices, the spatial coordinate is replaced by the discrete vertex set, and the state of the system at time $t$ is represented by the random vector
\begin{equation} \label{eq:stochastic-process}
    F(t;\boldsymbol{w})
    =
    \bigl(
        F_1(t;\boldsymbol{w}),\ldots,
        F_{N_{\mathcal V}}(t;\boldsymbol{w})
    \bigr)^\top
    \in \mathbb{R}^{N_{\mathcal V}},
\end{equation}
where $F_i(t;\boldsymbol{w})$ denotes the random state associated with the vertex $v_i$. In this sense, a stochastic process on a graph provides a finite-dimensional counterpart of a spatially distributed stochastic process, with the spatially indexed random field replaced by a collection of random variables indexed by the graph vertices.

Similarly, the spatial differential operator $\mathcal{L}$ that appears in an SPDE is replaced by a matrix-valued graph operator $\boldsymbol{L}(\boldsymbol{w})$, which encodes interactions between vertices. The graph Laplacian provides a discrete counterpart of the negative Laplace operator, while functions of the graph Laplacian allow more general differential operators, such as the Whittle--Matérn operator introduced previously, to be transferred to the graph setting. Reaction terms carry over by acting locally on the vertex states, while stochastic forcing is represented by a finite-dimensional Brownian motion. Thus, the principal components of the continuous SPDE have direct finite-dimensional counterparts on a graph.

Following this correspondence, we formulate the continuous-time graph process as the stochastic differential equation
\begin{equation}
    \label{eq:processR}
    dF(t;\boldsymbol{w})
    =
    -\boldsymbol{L}(\boldsymbol{w})F(t;\boldsymbol{w})\,dt
    +
    \beta\mathcal{R}\bigl(F(t;\boldsymbol{w})\bigr)\,dt
    +
    \boldsymbol{\Gamma}\,d\xi(t).
\end{equation}
Here, $\boldsymbol{L}(\boldsymbol{w})\in
\mathbb{R}^{N_{\mathcal V}\times N_{\mathcal V}}$ governs interactions between vertices, $\mathcal{R}:\mathbb{R}^{N_{\mathcal V}}\to \mathbb{R}^{N_{\mathcal V}}$ describes local reaction dynamics, $\boldsymbol{\Gamma}\in \mathbb{R}^{N_{\mathcal V}\times N_{\mathcal V}}$ determines the directions and magnitude of the stochastic forcing, and $\xi(t)\in\mathbb{R}^{N_{\mathcal V}}$ is an $N_{\mathcal V}$-dimensional Brownian motion. Furthermore, $F(t;\boldsymbol{w})\in\mathbb{R}^{N_{\mathcal V}}$ denotes the random state of the graph at time $t$. Conditional on a realization $W=\boldsymbol{w}$ of the graph weights, \eqref{eq:processR} therefore defines an $\mathbb{R}^{N_{\mathcal V}}$-valued stochastic process whose evolution is determined jointly by the graph structure, the reaction dynamics, and the Brownian forcing.

The stochastic differential equation \eqref{eq:processR} is understood through its integral representation
\begin{equation}
    \label{eq:process-integral}
    F(t;\boldsymbol{w})
    =
    F(0)
    +
    \int_0^t
        -\boldsymbol{L}(\boldsymbol{w})F(s;\boldsymbol{w})\,ds
    +
    \int_0^t
        \beta\mathcal{R}\bigl(F(s;\boldsymbol{w})\bigr)\,ds
    +
    \boldsymbol{\Gamma}
    \int_0^t d\xi(s),
\end{equation}
where the last term is understood as an It\^o stochastic integral. Here, $\xi(t)$ is an $N_{\mathcal V}$-dimensional Brownian motion with independent increments satisfying
\begin{equation}
    \xi(t+\Delta t)-\xi(t)
    \sim
    \mathcal{N}
    \left(
        0,
        \Delta t\,\boldsymbol{I}_{N_{\mathcal V}}
    \right).
\end{equation}
The matrix
$\boldsymbol{\Gamma}\in
\mathbb{R}^{N_{\mathcal V}\times N_{\mathcal V}}$ determines how the Brownian forcing acts on the vertex states and induces the instantaneous covariance
$\boldsymbol{\Gamma}\boldsymbol{\Gamma}^{\top}$.

For numerical simulation, we introduce a uniform temporal discretization
\begin{equation}
    0=t_0<t_1<\cdots<t_{N_T}=T,
    \qquad
    t_i=i\Delta t.
\end{equation}
Over each time interval, the Brownian increment can be represented as
\begin{equation}
    \Delta\xi_i
    :=
    \xi(t_{i+1})-\xi(t_i)
    =
    \sqrt{\Delta t}\,\boldsymbol{\varepsilon}_i,
    \qquad
    \boldsymbol{\varepsilon}_i
    \sim
    \mathcal{N}
    \left(
        0,\boldsymbol{I}_{N_{\mathcal V}}
    \right),
\end{equation}
where the vectors $\boldsymbol{\varepsilon}_i$ are independent. A temporal discretization of \eqref{eq:processR} therefore, gives the recursive process
\begin{equation}
    \label{eq:process_recursive}
    F(t_i;\boldsymbol{w})
    =
    F(t_{i-1};\boldsymbol{w})
    -
    \Delta t\,
    \boldsymbol{L}(\boldsymbol{w})
    F(t_{i-1};\boldsymbol{w})
    +
    \Delta t\,
    \beta\mathcal{R}
    \bigl(
        F(t_{\ell};\boldsymbol{w})
    \bigr)
    +
    \sqrt{\Delta t}\,
    \boldsymbol{\Gamma}
    \boldsymbol{\varepsilon}_{i-1},
\end{equation}
for $i=1,\dots,N_T$, with $F(t_0;\boldsymbol{w})=F(0)$. Here, $\ell=i-1$ or $\ell=i$ depending on whether the reaction term is treated explicitly or implicitly; see \Cref{sec:time-discretization} and SM.

Next, we specify the graph operator $\boldsymbol{L}(\boldsymbol{w})$ appearing in \eqref{eq:processR}. In the continuous setting, spatial diffusion is typically described using the Laplace operator. On a weighted graph, the corresponding interactions between vertices can be represented through the graph Laplacian
\begin{equation}
    \label{eq:graph_laplace}
    \boldsymbol{L}(\boldsymbol{w})
    =
    \boldsymbol{D}(\boldsymbol{w})
    -
    \boldsymbol{A}(\boldsymbol{w}),
\end{equation}
where $\boldsymbol{A}(\boldsymbol{w})$ is the weighted adjacency matrix and $\boldsymbol{D}(\boldsymbol{w})$ is the diagonal weighted degree matrix, defined by
\begin{equation}
    \label{eq:graph-degree}
    \bigl[
        \boldsymbol{D}(\boldsymbol{w})
    \bigr]_{ii}
    =
    \sum_{j=1}^{N_{\mathcal V}}
    \bigl[
        \boldsymbol{A}(\boldsymbol{w})
    \bigr]_{ij}.
\end{equation}
The graph Laplacian, therefore, encodes the coupling between neighboring vertices, with the edge weights controlling the strength of these interactions. Small edge weights restrict interactions between the corresponding vertices, whereas large weights strengthen them.

Sometimes it proves convenient to use the normalized graph Laplacian
\begin{equation}
    \label{eq:normalized-laplacian}
    \widetilde{\boldsymbol{L}}
    =
    \boldsymbol{S}^{\top}
    \boldsymbol{L}
    \boldsymbol{S},
\end{equation}
where $\boldsymbol{S}$ is the diagonal matrix $[\boldsymbol{S}]_{ii}
    =
    {1}/{\sqrt{[\boldsymbol{D}]_{ii}}},$ 
    $i=1,\dots,N_{\mathcal V}$.
For structured graphs arising from spatial discretizations, graph Laplacians of the form \eqref{eq:graph_laplace} are closely related to standard discrete approximations of the negative Laplace operator. More generally, however, \eqref{eq:graph_laplace} does not require the vertices to be embedded in a Euclidean spatial domain. This allows the same diffusion-type construction to be extended to arbitrary weighted graph structures.

The choice of the graph operator and the temporal discretization jointly determine the numerical stability properties of the discrete process \eqref{eq:process_recursive}. In the numerical experiments considered in this work, the time step $\Delta t$ is chosen sufficiently small to ensure stable numerical behavior for the particular graph operators and reaction models under consideration.

The operator analogy between continuous spatial domains and graphs also allows the Whittle--Matérn construction introduced in \Cref{sec:diffterm} to be transferred to the graph setting. 
Using the correspondence between the negative Laplace operator $-\Delta$
and the graph Laplacian $\boldsymbol{L}(\boldsymbol{w})$, the
Whittle--Matérn construction introduced in
Section~\ref{sec:diffterm} can be transferred to the graph setting,
yielding
\begin{equation}
    \label{eq:graph-whittle-matern}
    \boldsymbol{L}_{\kappa,\alpha}(\boldsymbol{w})
    :=
    \left(
        \kappa^2\boldsymbol{I}
        +
        \boldsymbol{L}(\boldsymbol{w})
    \right)^{\alpha/2}.
\end{equation}
Since $\boldsymbol{L}(\boldsymbol{w})$ is symmetric, the fractional power in
\eqref{eq:graph-whittle-matern} can be defined through its eigendecomposition.
In particular, if
\begin{equation}
    \begin{aligned}
        \boldsymbol{L}(\boldsymbol{w})
        =
        \boldsymbol{V}
        \boldsymbol{\Lambda}
        \boldsymbol{V}^{\top}, \text{ then, }
        \qquad
        \boldsymbol{L}_{\kappa,\alpha}(\boldsymbol{w})
        =
        \boldsymbol{V}
        \left(
            \kappa^2\boldsymbol{I}
            +
            \boldsymbol{\Lambda}
        \right)^{\alpha/2}
        \boldsymbol{V}^{\top}.
    \end{aligned}
\end{equation}
Thus, the graph eigenvectors play a role analogous to spatial frequency modes in the continuous setting. Importantly, because the graph Laplacian depends on the edge weights $\boldsymbol{w}$, the corresponding Whittle--Matérn operator also depend on the unknown graph.

The matrix $\boldsymbol{\Gamma}$ provides an additional mechanism for controlling the stochastic forcing in \eqref{eq:processR}. The choice
$\boldsymbol{\Gamma}=\boldsymbol{I}$ corresponds to independent Brownian forcing of equal magnitude at each vertex, while a general $\boldsymbol{\Gamma}$ introduces correlations or different forcing strengths across the vertex states. In particular, $\boldsymbol{\Gamma}$ may be chosen with respect to the spectral structure of the graph operator when specific graph-frequency components are to be emphasized or suppressed.

For fixed graph weights $\boldsymbol{w}$, the stochastic differential equation
\eqref{eq:processR} induces a probability law on the trajectories of
$F(t;\boldsymbol{w})$. We denote this conditional law by
$\mathcal{P}_{\boldsymbol{w}}$ and write
\begin{equation}
    \label{eq:conditional-process-law}
    (F\mid W=\boldsymbol{w})
    \sim
    \mathcal{P}_{\boldsymbol{w}}.
\end{equation}
The notation emphasizes an important distinction in the proposed model: conditioning on the graph weights fixes the graph operator $\boldsymbol{L}(\boldsymbol{w})$, but does not fix the realization of the process $F$. Even for a fixed graph, the process remains random due to the Brownian forcing in \eqref{eq:processR}. In the linear case, and under Gaussian initial conditions, the Brownian forcing induces a Gaussian stochastic process. In the presence of a nonlinear reaction term, however, the resulting process is generally non-Gaussian.

The resulting model, therefore, has a hierarchical structure. The graph weights determine the graph operator and hence the law of the stochastic process, while the observations are generated conditionally  on a realization of that process. Schematically,
\begin{equation}
    \label{eq:inverse-summary}
    \begin{aligned}
        W &\sim \pi_W,\\
        (F\mid W=\boldsymbol{w})
            &\sim \mathcal{P}_{\boldsymbol{w}},\\
        (Y\mid F=\boldsymbol{f})
            &\sim
            \mathcal{N}
            \left(
                \boldsymbol{f},
                \Sigma_{\mathrm{noise}}
            \right).
    \end{aligned}
\end{equation}
Equivalently, whenever the corresponding densities exist, the joint distribution admits the factorization
\begin{equation}
    \label{eq:hierarchical-factorization}
    \pi_{(Y,F,W)}
    (\boldsymbol{y},\boldsymbol{f},\boldsymbol{w})
    =
    \pi_{(Y\mid F=\boldsymbol{f})}(\boldsymbol{y})
    \,
    \pi_{(F\mid W=\boldsymbol{w})}(\boldsymbol{f})
    \,
    \pi_W(\boldsymbol{w}).
\end{equation}
The conditional distribution
$\pi_{(F\mid W=\boldsymbol{w})}$ implicitly accounts for the Brownian forcing driving \eqref{eq:processR}. Thus, although the Brownian motion does not appear explicitly in \eqref{eq:hierarchical-factorization}, its uncertainty is propagated through the conditional law
$\mathcal{P}_{\boldsymbol{w}}$.

For time-homogeneous dynamics, one may additionally consider a stationary regime. If, for fixed $\boldsymbol{w}$, the process admits an invariant probability measure, we denote this measure by $\mathcal{P}_{\boldsymbol{w}}^{\mathrm{stat}}$. When the drift is linear and the driving noise is Gaussian, the invariant measure, whenever it exists, is Gaussian. Denoting by $\boldsymbol{B}(\boldsymbol{w})\in\mathbb{R}^{N_{\nu}\times N_{\nu}}$ the matrix governing the linear drift, the dynamics can be written as
\begin{equation}
    dF(t)
    =
    \boldsymbol{B}(\boldsymbol{w})F(t)\,dt
    +
    \boldsymbol{\Gamma}\,d\xi(t),
\end{equation}
its stationary covariance matrix $\boldsymbol{C}_{\boldsymbol{w}}\in\mathbb{R}^{N_{\nu}\times N_{\nu}}$ satisfies the continuous Lyapunov equation
\begin{equation}
    \label{eq:stationary-lyapunov}
    \boldsymbol{B}(\boldsymbol{w})
    \boldsymbol{C}_{\boldsymbol{w}}
    +
    \boldsymbol{C}_{\boldsymbol{w}}
    \boldsymbol{B}(\boldsymbol{w})^{\top}
    +
    \boldsymbol{\Gamma}
    \boldsymbol{\Gamma}^{\top}
    =
    0.
\end{equation}
\subsection{Reaction terms}\label{sec:reaction-term}

We next specify the reaction term $\mathcal{R}$ appearing in \eqref{eq:processR}. In contrast to the graph diffusion operator, which couples the states of neighboring vertices, the reaction term acts locally on each vertex state. We consider a nonlinear reaction model motivated by Susceptible--Infectious--Susceptible (SIS) dynamics, i.e.

\begin{equation}
    \label{eq:R}
    \mathcal{R}(F)
    =
    (\psi-\gamma)F-\psi F^2,
\end{equation}
where, for the graph-valued state $F\in\mathbb{R}^{N_{\mathcal V}}$, the nonlinear operations are understood componentwise. Thus, $\bigl[\mathcal{R}(F)\bigr]_i=(\psi-\gamma)F_i-\psi F_i^2$, $i=1,\dots,N_{\mathcal V}$.

This model is a classical compartmental model for disease transmission involving susceptible and infectious populations \cite{Kuhl,Schenk2024}. Let $S$ and $I$ denote the susceptible and infectious fractions, respectively, and assume that the total population is constant and normalized to 1, such that $S+I=1$. The infectious fraction in the SIS model evolves as
\begin{equation}
    \frac{dI}{dt}
    =
    \psi S I-\gamma I,
\end{equation}
where $\psi>0$ denotes the transmission rate and $\gamma>0$ the recovery rate. Substituting $S=1-I$ gives $dI/dt=(\psi-\gamma)I-\psi I^2$, which coincides with \eqref{eq:R} when identifying $I=F$.

According to epidemiological interpretation, $F_i$ represents the proportion of infectious individuals associated with the vertex $v_i$ and therefore takes values in $[0,1]$. For $\psi>\gamma$, equivalently, when the basic reproduction number $R_0=\psi/\gamma$ satisfies $R_0>1$, the local SIS dynamics admits the nonzero endemic equilibrium
\begin{equation}
    \label{eq:endemic-equilibrium}
    F_0
    =
    1-\frac{\gamma}{\psi}
    =
    \frac{\psi-\gamma}{\psi}.
\end{equation}
The corresponding susceptible fraction is $S_0=\gamma/\psi$. In this work, we fix $\gamma=1$ and $\psi=2.5$.

Although \eqref{eq:R} is motivated by the SIS model, the quadratic reaction term is not restricted to an epidemiological interpretation. More generally, nonlinear reaction terms of this form can describe state-dependent local production or dissipation mechanisms in reaction--diffusion systems. For example, when $F$ represents temperature, $\mathcal{R}(F)$ may be interpreted as a nonlinear heat source or sink. Thus, the SIS model provides a physically interpretable example of the broader nonlinear reaction--diffusion framework considered here.

For comparison, we also consider a linearized reaction model, with its derivation and numerical experiments provided in the SM.
\subsection{Time discretization}\label{sec:time-discretization}
We next describe the temporal discretization of the stochastic graph process for the nonlinear reaction model introduced above. The latter leads to a nonlinear system when treated implicitly.

\paragraph{Implicit discretization of the nonlinear model}
For the nonlinear reaction model, we employ an implicit Euler--Maruyama discretization of \eqref{eq:processR}, in which both the graph interaction and reaction terms are evaluated at the new time point $t_i$. This gives
\begin{equation}
    \label{eq:nonlinear-implicit}
    F(t_i;\boldsymbol{w})
    =
    F(t_{i-1};\boldsymbol{w})
    -
    \Delta t\,\boldsymbol{L}_{\kappa,\alpha}(\boldsymbol{w})F(t_i;\boldsymbol{w})
    +
    \Delta t\,\beta\mathcal{R}\bigl(F(t_i;\boldsymbol{w})\bigr)
    +
    \sqrt{\Delta t}\,\boldsymbol{\Gamma}\boldsymbol{\varepsilon}_{i-1},
\end{equation}
where $\boldsymbol{\varepsilon}_{i-1}\sim\mathcal{N}(0,\boldsymbol{I}_{N_{\mathcal V}})$ is a standard Gaussian random vector. Rearranging \eqref{eq:nonlinear-implicit} yields
\begin{equation}
    \label{eq:nonlinear-implicit-system}
    \left(
        \boldsymbol{I}_{N_{\mathcal V}}
        +
        \Delta t\,\boldsymbol{L}_{\kappa,\alpha}(\boldsymbol{w})
    \right)
    F(t_i;\boldsymbol{w})
    -
    \Delta t\,\beta\mathcal{R}\bigl(F(t_i;\boldsymbol{w})\bigr)
    =
    F(t_{i-1};\boldsymbol{w})
    +
    \sqrt{\Delta t}\,\boldsymbol{\Gamma}\boldsymbol{\varepsilon}_{i-1}.
\end{equation}

This nonlinear system is solved at each time step using Newton's method with a convergence tolerance of $10^{-6}$ and a maximum of $20$ iterations. To improve robustness, we use a line-search procedure with at most $10$ steps per Newton iteration. For a fixed time step $t_i$, define the residual
\begin{equation}
    \mathcal{G}_i(F)
    =
    \left(
        \boldsymbol{I}_{N_{\mathcal V}}
        +
        \Delta t\,\boldsymbol{L}_{\kappa,\alpha}(\boldsymbol{w})
    \right)F
    -
    \Delta t\,\beta\mathcal{R}(F)
    -
    F(t_{i-1};\boldsymbol{w})
    -
    \sqrt{\Delta t}\,\boldsymbol{\Gamma}\boldsymbol{\varepsilon}_{i-1}.
\end{equation}
The corresponding Jacobian is
\begin{equation}
    \label{eq:nonlinear-jacobian}
    J_i(F)
    =
    \boldsymbol{I}_{N_{\mathcal V}}
    +
    \Delta t\,\boldsymbol{L}_{\kappa,\alpha}(\boldsymbol{w})
    -
    \Delta t\,\beta D\mathcal{R}(F),
\end{equation}
where $D\mathcal{R}(F)$ denotes the Jacobian of the reaction term. For the componentwise nonlinear reaction term in \eqref{eq:R}, this matrix is diagonal and is given by
\begin{equation}
    D\mathcal{R}(F)
    =
    \operatorname{diag}
    \left(
        (\psi-\gamma)-2\psi F_1,
        \ldots,
        (\psi-\gamma)-2\psi F_{N_{\mathcal V}}
    \right).
\end{equation}
\section{The B-GRASP framework}\label{sec:BGRASP}
\begin{figure}[t!]
    \centering
    \resizebox{\textwidth}{!}{%
    \begin{tikzpicture}[
        node distance=0.5cm,
        every node/.style={
            draw,
            rounded corners,
            align=center,
            text width=3.2cm,
            minimum height=2.5cm,
            font=\small
        },
        arrow/.style={
            ->,
            thick,
            >=Stealth
        }
    ]

    \node (prior) {
        \textbf{Define prior distribution}\\[2pt]
        Latent graph variables
    };

    \node (graph) [right=of prior] {
        \textbf{Construct weighted graph}\\[2pt]
        Stochastic forcing \& forward computation\\
        Graph weights \& graph Laplacian
    };

    \node (forward) [right=of graph] {
        \textbf{Solve stochastic graph model}\\[2pt]
        Stochastic graph dynamics with diffusion and reaction
    };

    \node (likelihood) [right=of forward] {
        \textbf{Construct observation likelihood}\\[2pt]
        Predicted vs.\ observed vertex states
    };

    \node (posterior) [right=of likelihood] {
        \textbf{Posterior inference \& uncertainty quantification}\\[2pt]
        Infer graph weights and quantify their uncertainty
    };

    \draw[arrow] (prior) -- (graph);
    \draw[arrow] (graph) -- (forward);
    \draw[arrow] (forward) -- (likelihood);
    \draw[arrow] (likelihood) -- (posterior);

    \end{tikzpicture}%
    }

    \caption{
        Computational dependency structure of B-GRASP, from latent graph variables and stochastic forcing to the likelihood, posterior, and uncertainty quantification of graph weights.
    }
    \label{fig:flowchart}
\end{figure}
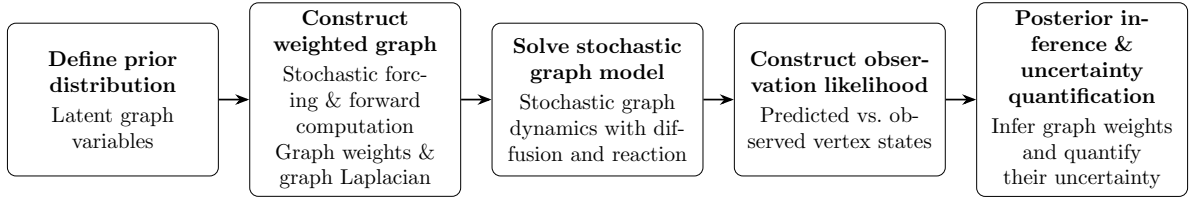

Having introduced the Bayesian inverse formulation and stochastic graph dynamics in \Cref{sec:invgraphprob} and \Cref{sec:graph-processes}, respectively, we now bring these components together to formulate the B-GRASP framework for uncertainty quantification of graph weights. B-GRASP combines latent representations of the graph weights and stochastic forcing with the graph forward model and observation likelihood to construct the posterior distribution. The resulting computational dependency structure is summarized in \Cref{fig:flowchart}.

Starting from a graph $\mathcal{G}=(\mathcal{V},\mathcal{E})$ with fixed vertices and edge connectivity, we assign a prior distribution to latent variables $X$ parameterizing the unknown graph weights, from which the weights $W=T(X)$ and graph Laplacian $\boldsymbol{L}(W)$ are constructed. In the time-dependent setting, latent Brownian variables $Z$ additionally determine a realization of the stochastic forcing that, together with the graph Laplacian, propagates the graph process according to the time discretization introduced previously. The resulting graph states are compared with observations through an additive Gaussian measurement model, yielding a posterior distribution over the latent graph variables and stochastic forcing from which uncertainty in the graph weights can be quantified.

\subsection{Prior distribution}
Let $X=(X_1,\ldots,X_{N_{\mathcal E}})^\top\in\mathbb{R}^{N_{\mathcal E}}$ denote the latent random vector associated with the $N_{\mathcal E}$ graph edges. In this work, we assign independent standard Gaussian priors to these variables,
\begin{equation}
X_j \overset{\mathrm{i.i.d.}}{\sim} \mathcal{N}(0,1),
\qquad
j=1,\ldots,N_{\mathcal E},
\end{equation}
or, equivalently, $X\sim\mathcal{N}\left(\boldsymbol{0},\boldsymbol{I}_{N_{\mathcal E}} \right)$.
The corresponding prior density is
\begin{equation}
\label{eq:latent-weight-prior}
\pi_X(\boldsymbol{x})
\propto
\exp\left(
-\frac{1}{2}\|\boldsymbol{x}\|_2^2
\right).
\end{equation}

Since the graph weights are required to be positive, the unconstrained Gaussian variables are mapped to the physical edge weights through a componentwise transformation $\mathcal{T}:\mathbb{R}^{N_{\mathcal E}}\rightarrow\mathbb{R}_{>0}^{N_{\mathcal E}}$, given by
\begin{equation}
\label{eq:weight-transformation}
W=\mathcal{T}(X),
\qquad
w_j
=
w_{\mathrm{mean}}
+
w_{\mathrm{scale}}\exp(x_j),
\qquad
j=1,\ldots,N_{\mathcal E}.
\end{equation}

For a realization $\boldsymbol{x}$ of $X$, the transformed weights $\boldsymbol{w}=\mathcal{T}(\boldsymbol{x})$ are assigned to the edges according to the fixed ordering of $\mathcal{E}$ and used to construct the adjacency matrix $\boldsymbol{A}(\boldsymbol{w})$ and the corresponding graph Laplacian $\boldsymbol{L}(\boldsymbol{w})$ according to \eqref{eq:graph_laplace}. When required, the normalized graph Laplacian is constructed according to \eqref{eq:normalized-laplacian}. Consequently, each realization of $X$ determines a realization of the graph weights and, hence, of the graph operator that governs the dynamics. This dependence of $\boldsymbol{L}$ on the unknown weights provides the mechanism through which uncertainty in the graph structure is propagated to the graph process.

\subsection{Stochastic forcing and forward computation}

In the time-dependent model, the graph state remains stochastic even after conditioning on the graph weights because of the Brownian forcing in \eqref{eq:processR}. Computationally, Brownian increments are represented by independent standard Gaussian random vectors $Z_i\in\mathbb{R}^{N_{\mathcal V}}$, with $Z_i\overset{\mathrm{i.i.d.}}{\sim}\mathcal{N}\left(\boldsymbol{0},\boldsymbol{I}_{N_{\mathcal V}}\right)$, $i=0,\ldots,N_T-1$, such that $\xi(t_{i+1})-\xi(t_i)=\sqrt{\Delta t}\,Z_i$.
We collect these variables into $Z=(Z_0,\ldots,Z_{N_T-1})$. Their joint density is therefore
\begin{equation}
\label{eq:brownian-latent-prior}
\pi_Z(\boldsymbol{z})
\propto
\exp\left(
-\frac{1}{2}
\sum_{i=0}^{N_T-1}
|\boldsymbol{z}_i|_2^2
\right).
\end{equation}

For fixed realizations $X=\boldsymbol{x}$ and $Z=\boldsymbol{z}$, the forward computation is deterministic. First, the physical graph weights $\boldsymbol{w}=\mathcal{T}(\boldsymbol{x})$ and the corresponding graph Laplacian $\boldsymbol{L}(\boldsymbol{w})$ are constructed. The graph process is then propagated from its initial state using the time-discretization scheme introduced in the previous section, with the stochastic increments given by $\sqrt{\Delta t}\,\boldsymbol{\Gamma}\boldsymbol{z}_i$. For the computational formulation, we introduce the forward map $\mathcal{F}$, which maps realizations of the latent graph variables $X$ and Brownian variables $Z$ to the corresponding graph states at the measurement times. For fixed $\boldsymbol{x}$ and $\boldsymbol{z}$, we define
\begin{equation}
    \label{eq:computational-forward-map}
    \mathcal{F}(\boldsymbol{x},\boldsymbol{z})
    :=
    \left(
        F(t_1;\mathcal{T}(\boldsymbol{x}),\boldsymbol{z}),
        \ldots,
        F(t_{N_T};\mathcal{T}(\boldsymbol{x}),\boldsymbol{z})
    \right),
\end{equation}
where the additional dependence on $\boldsymbol{z}$ indicates the realization of the Brownian increments used to generate the trajectory.

\subsection{Likelihood}

We assume that the states of the graph are measured with additive Gaussian noise according to
\begin{equation}
    \label{eq:computational-observation-model}
    Y
    =
    \mathcal{F}(X,Z)
    +
    E,
    \qquad
    E
    \sim
    \mathcal{N}
    \left(
        \boldsymbol{0},
        \boldsymbol{\Sigma}_{\mathrm{noise}}
    \right),
\end{equation}
where $E$ represents measurement noise and is independent of $X$ and $Z$. Conditional on $X=\boldsymbol{x}$ and $Z=\boldsymbol{z}$, the observations therefore satisfy $
    (Y\mid X=\boldsymbol{x},Z=\boldsymbol{z})
    \sim
    \mathcal{N}
    \left(
        \mathcal{F}(\boldsymbol{x},\boldsymbol{z}),
        \boldsymbol{\Sigma}_{\mathrm{noise}}
    \right)$.
The corresponding likelihood density is
\begin{equation}
    \label{eq:computational-likelihood}
    \pi_{(Y\mid X=\boldsymbol{x},Z=\boldsymbol{z})}(\boldsymbol{y})
    \propto
    \exp\left(
        -\frac{1}{2}
        \left[
            \boldsymbol{y}
            -
            \mathcal{F}(\boldsymbol{x},\boldsymbol{z})
        \right]^\top
        \boldsymbol{\Sigma}_{\mathrm{noise}}^{-1}
        \left[
            \boldsymbol{y}
            -
            \mathcal{F}(\boldsymbol{x},\boldsymbol{z})
        \right]
    \right),
\end{equation}
where $N_y$ denotes the total number of scalar observations.

\subsection{Posterior distribution}

The latent graph variables $X$ and the variables $Z$ representing the Brownian increments are assumed to be independent a priori. Their joint prior density therefore factorizes as
\begin{equation}
    \pi_{(X,Z)}(\boldsymbol{x},\boldsymbol{z})
    =
    \pi_X(\boldsymbol{x})\,
    \pi_Z(\boldsymbol{z}),
\end{equation}
where the densities $\pi_X$ and $\pi_Z$ are given in \eqref{eq:latent-weight-prior} and \eqref{eq:brownian-latent-prior}, respectively. Combining this prior with the likelihood in \eqref{eq:computational-likelihood}, Bayes' rule gives the joint posterior distribution
\begin{equation}
    \label{eq:computational-posterior}
    \pi_{(X,Z\mid Y=\boldsymbol{y})}(\boldsymbol{x},\boldsymbol{z})
    \propto
    \pi_{(Y\mid X=\boldsymbol{x},Z=\boldsymbol{z})}(\boldsymbol{y})\,
    \pi_X(\boldsymbol{x})\,
    \pi_Z(\boldsymbol{z}).
\end{equation}
The posterior in \eqref{eq:computational-posterior} therefore characterizes the joint uncertainty in the latent graph parameters and the realization of the stochastic forcing conditional on the observed graph states. The corresponding uncertainty in the physical graph weights is obtained through the transformation $W=\mathcal{T}(X)$ in \eqref{eq:weight-transformation}.

\paragraph{Posterior point estimates and sampling}

Once the posterior distribution in \eqref{eq:computational-posterior} has been defined, it may be summarized through point estimates or explored more fully through posterior sampling. A common point estimate is the posterior mean, while the MAP estimate identifies the point of highest posterior density. More generally, posterior moments such as the mean and variance can be written as expectations with respect to the posterior distribution. Since these expectations are generally not available in closed form, we approximate them using ergodic averages of samples generated from the posterior.

Let $\{(\boldsymbol{x}^{(m)},\boldsymbol{z}^{(m)})\}_{m=1}^{M}$ denote samples generated by a Markov chain \cite{kaipio2005statistical} with invariant distribution $\pi_{(X,Z\mid Y=\boldsymbol{y})}$, where $M$ is the number of retained samples. For a posterior quantity of interest $g(X,Z)$, its expectation can be approximated by the corresponding ergodic average
\begin{equation} \label{eq:ergodic}
    \mathbb{E}
    \left[
        g(X,Z)
        \mid
        Y=\boldsymbol{y}
    \right]
    \approx
    \frac{1}{M}
    \sum_{m=1}^{M}
    g\left(
        \boldsymbol{x}^{(m)},
        \boldsymbol{z}^{(m)}
    \right).
\end{equation}
In particular, posterior means and variances of the latent variables, graph weights, or derived graph quantities can be estimated directly from these samples. 

\begin{figure}[htb]
    \centering
    \resizebox{0.95\textwidth}{!}{%
    \begin{tikzpicture}[
        node distance=1.8cm and 2.2cm,
        every node/.style={font=\small},
        block/.style={
            draw,
            rounded corners,
            minimum height=0.9cm,
            minimum width=2.1cm,
            align=center
        },
        arrow/.style={->, thick}
    ]

        \node[block] (x) {$\boldsymbol{x}$};
        \node[block, right=of x] (w) {$\boldsymbol{w}=\mathcal{T}(\boldsymbol{x})$};
        \node[block, right=of w] (L) {$\boldsymbol{L}_{\kappa,\alpha}(\boldsymbol{w})$};

        \node[block, below=of L] (F) {$\mathcal{F}(\boldsymbol{x},\boldsymbol{z})$};
        \node[block, left=of F] (z) {$\boldsymbol{z}$};

        \node[block, right=of F] (like)
        {$\pi_{(Y\mid X=\boldsymbol{x},Z=\boldsymbol{z})}(\boldsymbol{y})$};

        \node[block, right=of like] (post)
        {$\pi_{(X,Z\mid Y=\boldsymbol{y})}(\boldsymbol{x},\boldsymbol{z})$};

        \draw[arrow] (x) -- (w);
        \draw[arrow] (w) -- (L);
        \draw[arrow] (L) -- (F);
        \draw[arrow] (z) -- (F);
        \draw[arrow] (F) -- (like);
        \draw[arrow] (like) -- (post);

        \draw[arrow, bend left=25]
        (x) to node[above] {$\pi_X$} (post);

        \draw[arrow, bend right=25]
        (z) to node[below] {$\pi_Z$} (post);

    \end{tikzpicture}%
    }
    \caption{Dependency structure of the computational posterior. The latent graph variables $\boldsymbol{x}$ determine the graph weights and graph Laplacian, while $\boldsymbol{z}$ determines the realization of the stochastic forcing. Both enter the computational forward map $\mathcal{F}$ and subsequently the likelihood and posterior distribution.}
    \label{fig:gradient-dependencies}
\end{figure}
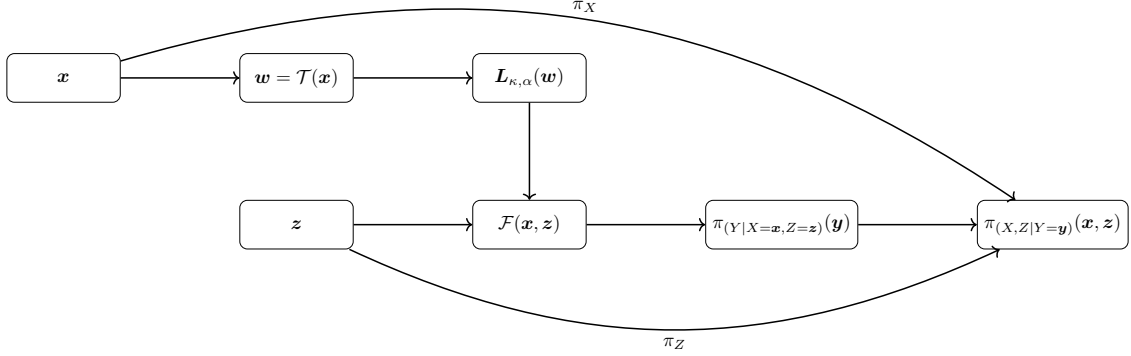

To explore the posterior, we use NUTS, an adaptive variant of Hamiltonian Monte Carlo (HMC) that utilizes gradient information for efficient posterior exploration. A central requirement of HMC and NUTS is the gradient of the log-posterior with respect to the sampled variables $\boldsymbol{x}$ and $\boldsymbol{z}$. The dependencies entering this gradient are summarized in \Cref{fig:gradient-dependencies}. 

All maps appearing in \Cref{fig:gradient-dependencies} are differentiable under the assumptions used here, and the required gradients of the negative log-posterior can therefore be evaluated by repeated application of the chain rule. In particular, differentiation with respect to $\boldsymbol{x}$ propagates through the sequence
\begin{equation}
    \boldsymbol{x}
    \longmapsto
    \mathcal{T}(\boldsymbol{x})
    \longmapsto
    \boldsymbol{L}_{\kappa,\alpha}(\mathcal{T}(\boldsymbol{x}))
    \longmapsto
    \mathcal{F}(\boldsymbol{x},\boldsymbol{z})
    \longmapsto
    \log\pi_{(X,Z\mid Y=\boldsymbol{y})},
\end{equation}
whereas differentiation with respect to $\boldsymbol{z}$ propagates directly through the stochastic forcing and the forward map $\mathcal{F}$.

For the unnormalized graph Laplacian, the dependence on an individual edge weight has a particularly simple form. Let $e=(v_i,v_j)\in\mathcal{E}$ denote an edge with weight $w_e$. Then
\begin{equation}
    \label{eq:laplacian-weight-derivative}
    \frac{\partial\boldsymbol{L}}{\partial w_e}
    =
    (\boldsymbol{e}_i-\boldsymbol{e}_j)
    (\boldsymbol{e}_i-\boldsymbol{e}_j)^\top,
\end{equation}
where $\boldsymbol{e}_i\in\mathbb{R}^{N_{\mathcal V}}$ denotes the $i$th Euclidean basis vector. Equivalently, $\partial\boldsymbol{L}/\partial w_e$ has entries $+1$ at $(i,i)$ and $(j,j)$, entries $-1$ at $(i,j)$ and $(j,i)$, and zeros elsewhere.

The computational procedure for constructing the likelihood from the unknown graph parameters and the stochastic forcing is summarized in \Cref{alg:likelihood-construction}.

\begin{algorithm}[t]
\caption{Construction of the likelihood from a weighted graph}
\label{alg:likelihood-construction}
\begin{algorithmic}[1]

\Require Graph $\mathcal{G}=(\mathcal{V},\mathcal{E})$, latent variables $\boldsymbol{x}$ and $\boldsymbol{z}$, observations $\boldsymbol{y}$

\State Compute the graph weights $\boldsymbol{w}=\mathcal{T}(\boldsymbol{x})$ according to \eqref{eq:weight-transformation}.

\State Construct the graph Laplacian $\boldsymbol{L}(\boldsymbol{w})$ or $\widetilde{\boldsymbol{L}}$ according to \eqref{eq:graph_laplace} according to \eqref{eq:normalized-laplacian}.

\State Solve the graph process \eqref{eq:processR} with stochastic increments $\sqrt{\Delta t}\,\boldsymbol{\Gamma}\boldsymbol{z}_i$.

\State Collect the graph states at the measurement times according to \eqref{eq:computational-forward-map}.

\State Evaluate the likelihood $\pi_{(Y\mid X=\boldsymbol{x},Z=\boldsymbol{z})}(\boldsymbol{y})$ according to \eqref{eq:computational-likelihood}.

\end{algorithmic}
\end{algorithm}

\section{Results}\label{sec:res}
In this section, we evaluate the proposed Bayesian inverse graph methodology on a sequence of test problems of increasing complexity. We first consider a one-dimensional heat-propagation problem represented by a linear tree graph with prescribed heat sources and sinks. This example provides a simple setting in which to compare the graph-based formulation with a classical numerical PDE model and to illustrate the interpretation of graph weights as diffusion parameters. We then consider a hypothetical contamination or epidemic-exposure model defined on a graph of the contiguous United States, which is used to investigate the proposed methodology in both stationary and time-dependent settings. Finally, we consider a graph-based model of the spread of COVID-19 across the United States using observations collected during the pandemic. In this application, the objective is to infer the graph connectivity parameters from the observed state-level dynamics, and thereby identify edges associated with stronger transmission between connected states. All Python code used to generate the simulations in this paper is available online\footnote{\url{https://github.com/babakmaboudi/B-GRASP}} \cite{maboudiafkham2026bgrasp}

\subsection{1D heat problem with inhomogeneous conductivity}
We first consider the inverse problem of inferring an inhomogeneous conductivity field in a one-dimensional heat transfer model. This classical PDE-based inverse problem provides a transparent connection between spatially varying conductivity and graph edge weights and serves to illustrate the relationship between the PDE- and graph-based formulations.

\paragraph{Comparison with a finite-difference discretization in divergence form}
To connect the graph diffusion operator with a classical PDE discretization, consider the one-dimensional weighted negative Laplacian in divergence form, $\mathcal{L}_{w} f:=-d/dx\left(w(x)df/dx\right)$.
On a uniform grid $x_i=i\Delta x$, let $F_i\approx f(x_i)$ and represent the conductivity at the interfaces by $w_{i+1/2}=w(x_i+\Delta x/2)$. A standard second-order FD discretization gives
\begin{equation}
    \label{eq:FD-Laplacian}
    (\boldsymbol{L}_{w}\boldsymbol{F})_i
    =
    -\frac{1}{\Delta x^2}
    \left[
        w_{i+1/2}(F_{i+1}-F_i)
        -
        w_{i-1/2}(F_i-F_{i-1})
    \right].
\end{equation}
We impose homogeneous Neumann boundary conditions at both endpoints. 

For $\Delta x=1$, this discretization coincides with the action of the unnormalized weighted graph Laplacian $\boldsymbol{L}(\boldsymbol{w})$, with the interface conductivities $w_{i+1/2}$ identified with the graph edge weights. The endpoint vertices provide the corresponding zero-flux boundary treatment. Further details on the graph–finite-difference correspondence are provided in the SM.
\subsubsection{Inverse graph heat problem}
\begin{figure}[t]
    \centering
    \begin{tikzpicture}[
        vertex/.style={circle, fill=black, inner sep=2.5pt},
        edge/.style={thick},
        every node/.style={font=\small}
    ]

        \node[vertex] (v1) at (0,0) {};
        \node[vertex] (v2) at (2.3,0) {};
        \node[vertex] (v3) at (4.6,0) {};
        \node (dots) at (6.6,0) {$\cdots$};
        \node[vertex] (vNm1) at (8.6,0) {};
        \node[vertex] (vN) at (10.9,0) {};

        \draw[edge] (v1) -- (v2);
        \draw[edge] (v2) -- (v3);
        \draw[edge] (v3) -- (5.6,0);
        \draw[edge] (7.6,0) -- (vNm1);
        \draw[edge] (vNm1) -- (vN);

        \node[above=6pt of v1] {$F_1(t;\boldsymbol{w})$};
        \node[above=6pt of v2] {$F_2(t;\boldsymbol{w})$};
        \node[above=6pt of v3] {$F_3(t;\boldsymbol{w})$};
        \node[above=6pt of vNm1] {$F_{N_{\mathcal V}-1}(t;\boldsymbol{w})$};
        \node[above=6pt of vN] {$F_{N_{\mathcal V}}(t;\boldsymbol{w})$};

        \node[below=18pt of v1] {$v_1$};
        \node[below=18pt of v2] {$v_2$};
        \node[below=18pt of v3] {$v_3$};
        \node[below=18pt of vNm1] {$v_{N_{\mathcal V}-1}$};
        \node[below=18pt of vN] {$v_{N_{\mathcal V}}$};

        \node[below] at (1.15,0) {$w_1=[\mathcal{T}(\boldsymbol{x})]_1$};
        \node[below] at (3.45,0) {$w_2=[\mathcal{T}(\boldsymbol{x})]_2$};
        \node[below] at (9.75,0) {$w_{N_{\mathcal E}}=[\mathcal{T}(\boldsymbol{x})]_{N_{\mathcal E}}$};

    \end{tikzpicture}

    \caption{Path graph $\mathcal{G}=(\mathcal{V},\mathcal{E})$ with $N_{\mathcal V}$ vertices and $N_{\mathcal E}=N_{\mathcal V}-1$ edges. The state $F_i(t;\boldsymbol{w})$ is associated with vertex $v_i$, while the edge weights $\boldsymbol{w}=\mathcal{T}(\boldsymbol{x})$ govern interactions between neighboring vertices.}
    \label{fig:graph-linear}
\end{figure}
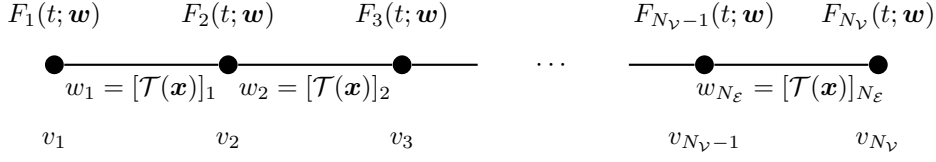

The one-dimensional inverse heat problem consists of inferring the conductivity of a heterogeneous one-dimensional medium (a rod) from temperature measurements. Let $\Omega=[0,L]$, $L>0$, denote the spatial domain of the rod. Heat transfer in an inhomogeneous medium is modeled by the stochastic heat equation
\begin{equation}
    \label{eq:1d-heat}
    df(x,t)
    =
    \frac{\partial}{\partial x}
    \left(
        w(x)\frac{\partial f(x,t)}{\partial x}
    \right)dt
    +
    \Gamma\,d\xi(x,t),
    \;\;
    f(x,0)=f_0(x),
    \;\;
    \frac{\partial f}{\partial x}\bigg|_{x=0}
    =
    \frac{\partial f}{\partial x}\bigg|_{x=L}
    =0.
\end{equation}

Here, $f(x,t)$ denotes temperature, $w(x)>0$ the spatially varying conductivity, $\xi$ the stochastic forcing, and $f_0$ the initial condition. Upon spatial discretization, the nodal temperatures are represented by $F_i(t)=f(x_i,t)$, while the interface conductivities are identified with the graph weights $\boldsymbol{w}=\mathcal{T}(\boldsymbol{x})=\exp(\boldsymbol{x})$, as illustrated in Figure~\ref{fig:graph-linear}. The Brownian increments are parameterized by independent standard Gaussian variables $\boldsymbol{z}_i$, as introduced previously. The inverse problem therefore consists of inferring $\boldsymbol{x}$ and $\boldsymbol{z}$ from noisy temperature observations.

We discretize the heat equation using the unnormalized graph Laplacian $\boldsymbol{L}(\boldsymbol{w})$ with $\Delta x=1$ and temporal discretization from \eqref{eq:process_recursive}. We choose a spatially correlated stochastic forcing by setting $\kappa^2=1$ and $\alpha=4$, and define the operator acting on the Gaussian increments as $\boldsymbol{\Gamma}(\boldsymbol{w}):=\boldsymbol{L}_{\kappa,\alpha}(\boldsymbol{w})^{-1}=\left(\boldsymbol{I}+\boldsymbol{L}(\boldsymbol{w})\right)^{-2}$. The resulting inverse problem follows the Bayesian formulation summarized in \eqref{eq:inverse-summary}, with the graph weights parameterized by $\boldsymbol{w}=\mathcal{T}(\boldsymbol{x})=\exp(\boldsymbol{x})$.

\begin{figure}[htb]
    \centering
    \includegraphics[width=0.75\linewidth, trim=10 6 7 9, clip]{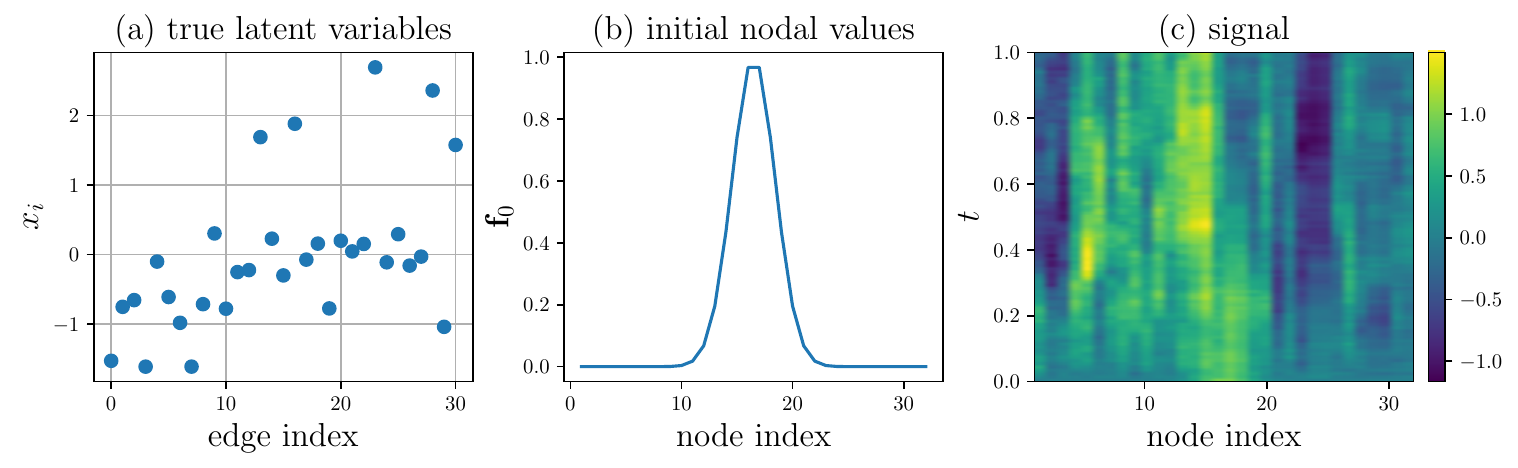}
    \caption{Signal used in the graph inverse heat problem (a) the true graph weights (b) initial condition for the heat problem (c) noise-free signal.}
    \label{fig:graphical_heat_signal}
\end{figure}

We choose $f_0(x)=\exp\left(-\left((x/L-0.5)/0.25\right)^2\right)$, use $\Delta t=0.01$ and $T=1$, and draw the ground-truth latent graph variables as $\boldsymbol{x}^{\mathrm{true}}\sim\mathcal{N}(\boldsymbol{0},\boldsymbol{I})$, giving $\boldsymbol{w}^{\mathrm{true}}=\exp(\boldsymbol{x}^{\mathrm{true}})$. The resulting weights, initial condition, and noise-free state trajectory are shown in \Cref{fig:graphical_heat_signal}. Observations are generated by adding independent Gaussian noise with $\sigma_{\mathrm{noise}}=0.01\|\boldsymbol{y}^{\mathrm{true}}\|_2$ and covariance $\boldsymbol{\Sigma}{\mathrm{noise}}=\sigma_{\mathrm{noise}}^2\boldsymbol{I}$.

Let $\mathcal{F}(\boldsymbol{x},\boldsymbol{z})$ denote the discrete stochastic heat-equation forward map evaluated at the measurement times. With independent standard Gaussian priors on $X$ and $Z$ and additive Gaussian observation noise, the log-posterior is, up to an additive constant,
\begin{equation}
    \label{eq:log-posterior-graphical-heat}
    \log \pi_{(X,Z\mid Y=\boldsymbol{y})}(\boldsymbol{x},\boldsymbol{z})
    =
    -\frac{1}{2}
    \left(
        \frac{
            \left\|
                \mathcal{F}(\boldsymbol{x},\boldsymbol{z})
                -
                \boldsymbol{y}
            \right\|_2^2
        }{
            \sigma_{\mathrm{noise}}^2
        }
        +
        \|\boldsymbol{x}\|_2^2
        +
        \|\boldsymbol{z}\|_2^2
    \right).
\end{equation}
We characterize this posterior using MAP estimation and NUTS as described in \Cref{sec:BGRASP}.
Gradients of the forward map with respect to $\boldsymbol{x}$ and $\boldsymbol{z}$ are computed using automatic differentiation in PyTorch \cite{paszke2019pytorch}. 
The MAP estimate is computed using the limited-memory Broyden–Fletcher–Goldfarb–Shanno (L-BFGS) algorithm \cite{liu1989limited}, while posterior sampling uses NUTS with 1000 warm-up iterations followed by 2000 sampling iterations. See the SM for more implementation and diagnostic details.

\begin{figure}[htb]
    \centering        \includegraphics[width=.7\linewidth, trim=8 6 5 5, clip]{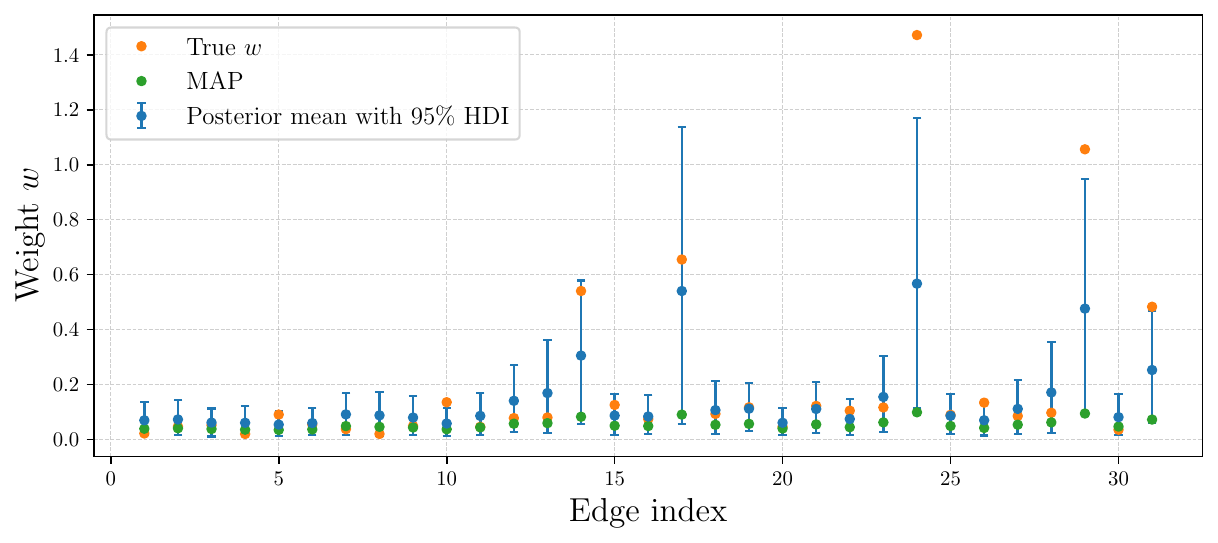}
        \caption{Posterior mean and MAP estimate with $95\%$ highest posterior density interval.}
        \label{fig:graphical_heat_estimation}
\end{figure}

\Cref{fig:graphical_heat_estimation} compares the posterior mean and MAP estimates with the ground-truth graph weights and reports the corresponding 95\% edge-wise highest density intervals (HDIs). For low to moderate conductivities, the posterior concentrates around the true values and both point estimates provide accurate reconstructions. For highly conductive edges, however, the point estimates tend to underestimate the true conductivity and posterior uncertainty increases substantially. Importantly, the 95\% HDIs extend beyond the point estimates and contain most extreme true conductivities, indicating that although the largest edge weights are weakly identified, the posterior captures the broader range compatible with the observations. This illustrates the value of Bayesian uncertainty quantification: information about highly conductive pathways obscured by point estimates remains visible in the posterior.

Because the Brownian variables are retained explicitly as latent variables, the framework also permits posterior reconstruction of the stochastic forcing; corresponding results are presented in the SM.
\subsection{Graphical heat problem}
In the previous section, we compared graph diffusion on a path graph with
the classical one-dimensional heat equation. We now consider a planar
graph representing geographical connectivity among the states of the
United States and investigate graph-weight inference under stationary and
non-stationary reaction--diffusion dynamics.

The stationary problem is driven by a persistent localized source and is
used to investigate how graph excitation affects edge identifiability.
The non-stationary problem instead uses time-series observations and
allows us to assess the additional information provided by the temporal
evolution of the graph state. We consider the nonlinear reaction model from \Cref{sec:reaction-term}; results for the linearized model for all cases are reported in the SM.

These synthetic examples illustrate the qualitative behavior of the proposed inference framework rather than represent realistic epidemiological scenarios. The non-stationary setting also provides a precursor to the COVID-19 application considered subsequently.

\subsubsection{Synthetic stationary graph reaction-diffusion problem in the United States} \label{sec:stationary-heat}
\begin{figure}
    \centering
    \includegraphics[width=1\linewidth, trim=5 5 5 5, clip]{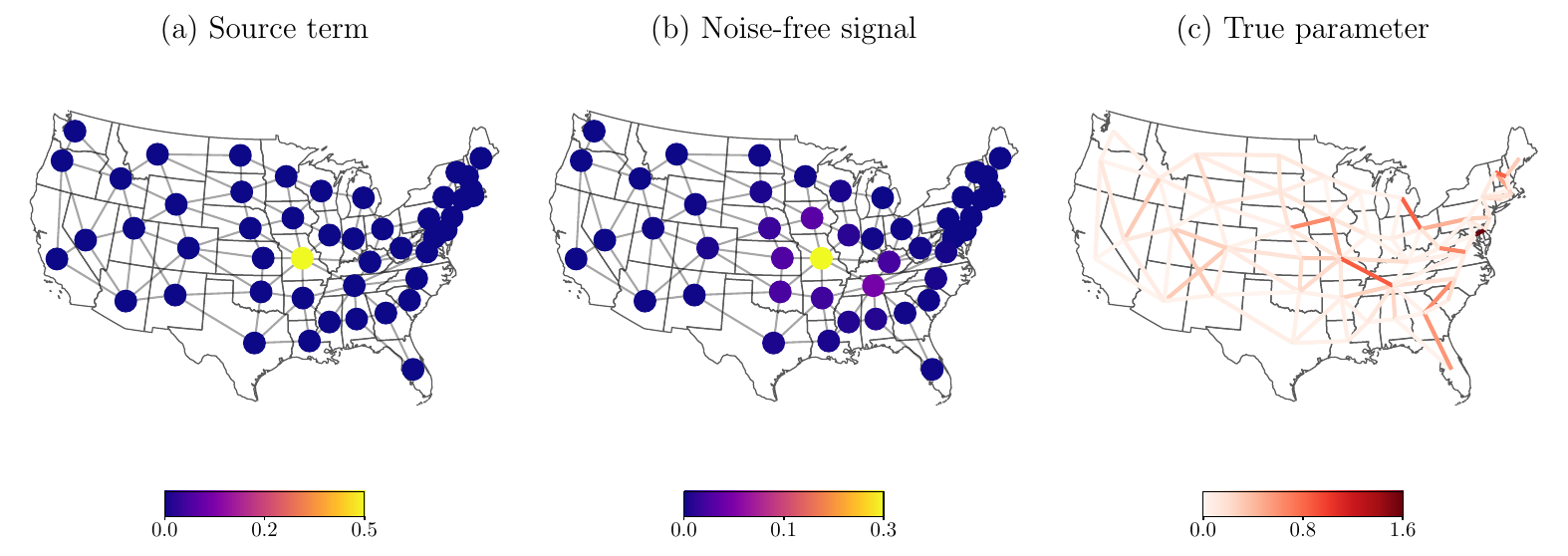}
    \caption{Data used in the steady state reaction-diffusion test problem. (a) localized source term (b) steady state solution with nonlinear reaction term (c) ground truth graph edge weights. }
    \label{fig:stationary_signal}
\end{figure}

Let $\mathcal{G}=(\mathcal{V},\mathcal{E})$ denote the US graph (see, e.g., \Cref{fig:stationary_signal}), consisting of $N_{\mathcal V}=51$ vertices representing the US states and $N_{\mathcal E}=110$ edges representing geographical connectivity between them. Alaska and Hawaii are retained as isolated vertices and are therefore dynamically uncoupled from the remaining graph. Their states are included in the computations but omitted from the visualization of the results.  

We collect the edge weights into a vector $\boldsymbol{w}\in\mathbb{R}_{>0}^{N_{\mathcal E}}$. From these weights, we construct the corresponding adjacency matrix $\boldsymbol{A}(\boldsymbol{w})$, degree matrix $\boldsymbol{D}(\boldsymbol{w})$, and diagonal scaling matrix $\boldsymbol{S}(\boldsymbol{w})=\boldsymbol{D}(\boldsymbol{w})^{-1/2}$. The symmetric normalized graph Laplacian $\widetilde{\boldsymbol{L}}(\boldsymbol{w})$ is then constructed according to \eqref{eq:normalized-laplacian}. 
For the numerical experiments, we set $\kappa^2=1$ and $\alpha=4$,
so that the graph Whittle--Matérn operator introduced in
\eqref{eq:graph-whittle-matern} becomes $
    \label{eq:wm-operator-us}
    \boldsymbol{L}_{\kappa,\alpha}(\boldsymbol{w})
    =
    \left(
        \boldsymbol{I}
        +
        \widetilde{\boldsymbol{L}}(\boldsymbol{w})
    \right)^2$.
The stationary graph reaction--diffusion equation, considered in this
section, is then
\begin{equation}
        \label{eq:stationary-heat-lin-nonlin}
    \boldsymbol{L}_{\kappa,\alpha}(\boldsymbol{w})F
    =
    \beta\mathcal{R}(F)
    +
    \boldsymbol{s},
\end{equation}
where $\boldsymbol{s}\in\mathbb{R}^{N_{\mathcal V}}$ is the source
term and $\mathcal{R}$ is the nonlinear reaction term, as introduced in \Cref{sec:reaction-term}. Here, we set the reaction weight to $\beta=10^{-2}$.

To generate the ground truth, we consider a persistent localized source
in Missouri,
$\boldsymbol{s}=(0,\ldots,2,\ldots,0)^{\mathsf T}$.
The ground-truth latent variables are drawn from the prescribed Gaussian
prior and mapped to positive edge weights following
$\boldsymbol{w}_{\mathrm{true}}
=\mathcal{T}(\boldsymbol{x}_{\mathrm{true}})
=\exp(\boldsymbol{x}_{\mathrm{true}})$.
Solving \eqref{eq:stationary-heat-lin-nonlin} with these weights and the
nonlinear reaction term yields the ground-truth stationary state
$\boldsymbol{F}_{\mathrm{true}}$.
\Cref{fig:stationary_signal} shows the source, stationary state, and ground-truth graph weights.

We generate noisy observations according to
\begin{equation}
    \label{eq:noisy-measurement}
    \boldsymbol{y}
    =
    \boldsymbol{F}_{\mathrm{true}}
    +
    \sigma_{\mathrm{noise}}\boldsymbol{e},
    \qquad
    \boldsymbol{e}
    \sim
    \mathcal{N}
    \left(
        \boldsymbol{0},
        \boldsymbol{I}_{N_{\mathcal V}}
    \right),
\end{equation}
where
$\sigma_{\mathrm{noise}}
=0.01\,\|\boldsymbol{F}_{\mathrm{true}}\|_2$.
As before, the positive graph weights are parameterized by
$\boldsymbol{w}=\exp(\boldsymbol{x})$, with a standard Gaussian prior
on $\boldsymbol{x}$. The posterior formulation and inference procedure follow the B-GRASP framework introduced in \Cref{sec:BGRASP}.

\begin{figure}
    \centering
    \includegraphics[width=1\linewidth, trim=10 6 7 8, clip]{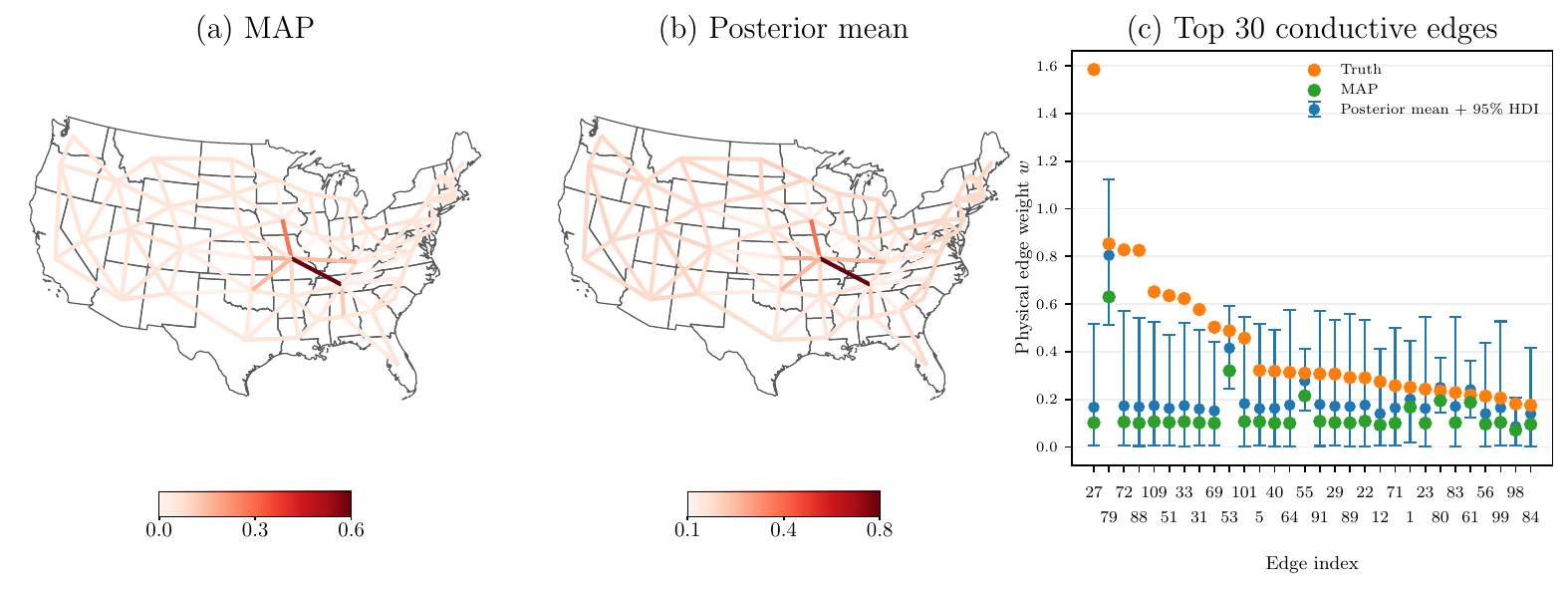}
    \caption{Estimation of the graph weights for the steady-state reaction diffusion model with non-linear reaction term. (a) MAP estimation of graph weights (b) posterior mean (c) posterior edge-wise standard deviation.}
    \label{fig:stationary_nonlinearreaction}
\end{figure}

\Cref{fig:stationary_nonlinearreaction} compares the posterior mean and
MAP estimates with the ground-truth graph weights and reports the
corresponding $95\%$ HDIs for the 30 most conductive edges.
The reconstruction reveals a clear dependence of edge identifiability
on graph excitation. Low-conductivity edges are generally recovered
accurately with narrow HDIs, while highly conductive edges near the
localized source are clearly identified and distinguished from the
low-conductivity background. Although the posterior mean and MAP
estimates underestimate some of the largest conductivities, the
corresponding HDIs reflect the greater uncertainty associated with these
edges.

In contrast, some highly conductive edges farther from the source
(e.g., edge 27) are only weakly informed by the observations, as shown
in \Cref{fig:stationary_nonlinearreaction}. 
As the signal diffuses away from the localized source, distant parts of the graph are less strongly affected, so variations in their conductivities have comparatively little influence on the observed stationary state. Thus, high conductivity
alone does not guarantee identifiability; an edge must also be
dynamically relevant under the prescribed excitation. Recovering the
complete weighted graph would therefore generally require richer
excitation.
\subsubsection{Synthetic non-stationary graph reaction-diffusion problem in the United States} \label{sec:nonstationary-heat}
We next consider the inference of graph edge weights from noisy
time-series observations of non-stationary graph reaction--diffusion
dynamics. We use the same graph structure and normalized graph Laplacian
as in \Cref{sec:stationary-heat}, allowing us to assess the additional
information provided by temporal observations relative to the stationary
problem.
The non-stationary graph reaction--diffusion equation is
\begin{equation}
    dF(t;\boldsymbol{w})
    =
    -\left(
        \boldsymbol{I}
        +
        \widetilde{\boldsymbol{L}}(\boldsymbol{w})
    \right)^2
    F(t;\boldsymbol{w})\,dt
    +
    \beta\mathcal{R}(F(t;\boldsymbol{w}))\,dt
    +
    \boldsymbol{\Gamma}\,d\xi(t),
\end{equation}
where $\mathcal{R}$ is the nonlinear reaction term introduced in
\Cref{sec:reaction-term} with $\beta=10^{-2}$ and the temporal
discretization from\eqref{eq:process_recursive}. The initial
graph state $F(0)$ is regarded as deterministic and known, and the
observations consist of nodal states recorded at each time step.

To generate the ground truth, we draw the initial graph state $F(0)$
componentwise from $\mathcal{U}(0,0.1)$ and generate the ground-truth
edge weights as in the stationary experiment. We then evolve the
nonlinear system using the temporal discretization in
\eqref{eq:process_recursive}, with $\Delta t=0.01$ and $T=5$, to obtain
the noise-free time series $\boldsymbol{F}_{\mathrm{true}}$.

Noisy observations follow
\eqref{eq:noisy-measurement}, with
$\sigma_{\mathrm{noise}}
=0.01\,\|\boldsymbol{F}_{\mathrm{true}}\|_2$.
As in the one-dimensional non-stationary problem, the Brownian variables
$\boldsymbol{z}$ are inferred jointly with the latent graph variables
$\boldsymbol{x}$, with $\boldsymbol{w}=\exp(\boldsymbol{x})$.
The posterior and inference procedure follow
\Cref{sec:BGRASP}. For this synthetic experiment, the nodal states are
restricted componentwise to $[0,1]$.

\begin{figure}[t!]
    \centering
    \includegraphics[width=1\linewidth, trim=10 8 7 8, clip]{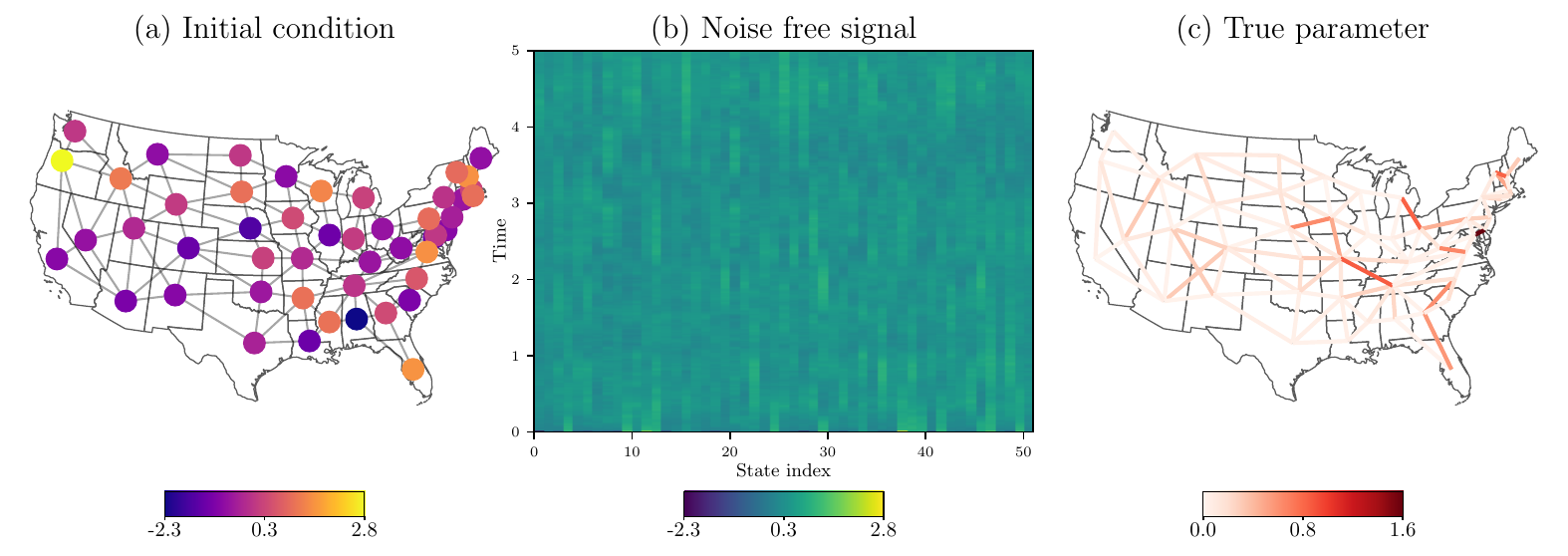}
    \caption{Data used for the non-stationary reaction-diffusion problem with nonlinear reaction term. (a) Initial condition (b) noise free signal (c) ground truth graph edge weights.}
    \label{fig:heat_signal}
\end{figure}

\begin{figure}[t!]
    \centering
    \includegraphics[width=1\linewidth, trim=10 6 7 8, clip]{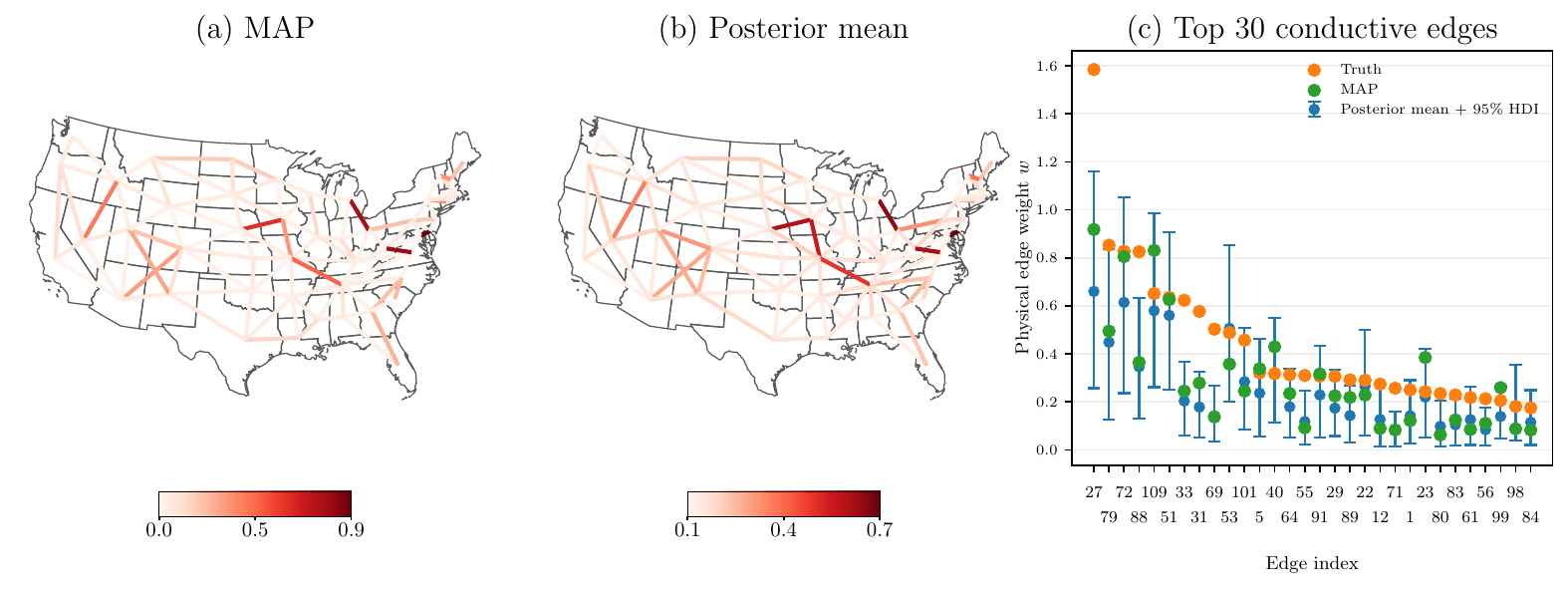}
    \caption{Estimation of the graph weights for the non-stationary reaction diffusion model with non-linear reaction term. (a) MAP estimation of graph weights (b) posterior mean (c) posterior edge-wise standard deviation.}
    \label{fig:heat_nonlinearreaction}
\end{figure}

\Cref{fig:heat_signal} shows the initial condition, noise-free trajectory, and ground-truth graph
weights.
The MAP estimate, posterior mean,
and the most conductive edges together with their $95\%$ HDIs are illustrated in \Cref{fig:heat_nonlinearreaction}. Many of
the dominant conductive pathways are identified by both the MAP and
posterior mean estimates. Compared with the corresponding stationary
nonlinear problem, the HDIs are generally narrower, indicating that the
time-resolved observations provide additional information for identifying
the edge conductivities. In particular, the most conductive true edge is
correctly identified as the dominant edge, and its magnitude is more
accurately represented.

Despite this improvement, some highly conductive edges remain outside
their corresponding $95\%$ HDIs. Consistent with the stationary
nonlinear problem, these edges appear to be only weakly excited by the
dynamics and therefore have limited influence on the observations.
Thus, temporal information improves inference for dynamically active
edges, while high conductivity alone remains insufficient for accurate
recovery when an edge is only weakly involved in the observed
transmission process.

\subsubsection{Nonstationary graph reaction–diffusion dynamics of COVID-19 in the United States} \label{sec:nonstationary-heat}
We next consider real-world observations using U.S. state-level cumulative COVID-19 case counts from The New York Times \cite{NYtimesCoviddata}, normalized by state population using the mappings from \cite{NikitinCovid}. The data comprise $N_{\mathcal V}=51$ graph vertices and $N_t=166$ observation times between January 21, 2020, and March 23, 2023.

The model and inference configuration follow the non-stationary synthetic
problem in \Cref{sec:nonstationary-heat}, except that no projection of the
graph state $F$ onto $[0,1]$ is applied and no artificial measurement
noise is added.

\begin{figure}[t!]
    \centering
    \includegraphics[width=1\linewidth, trim=10 8 7 8, clip]
    {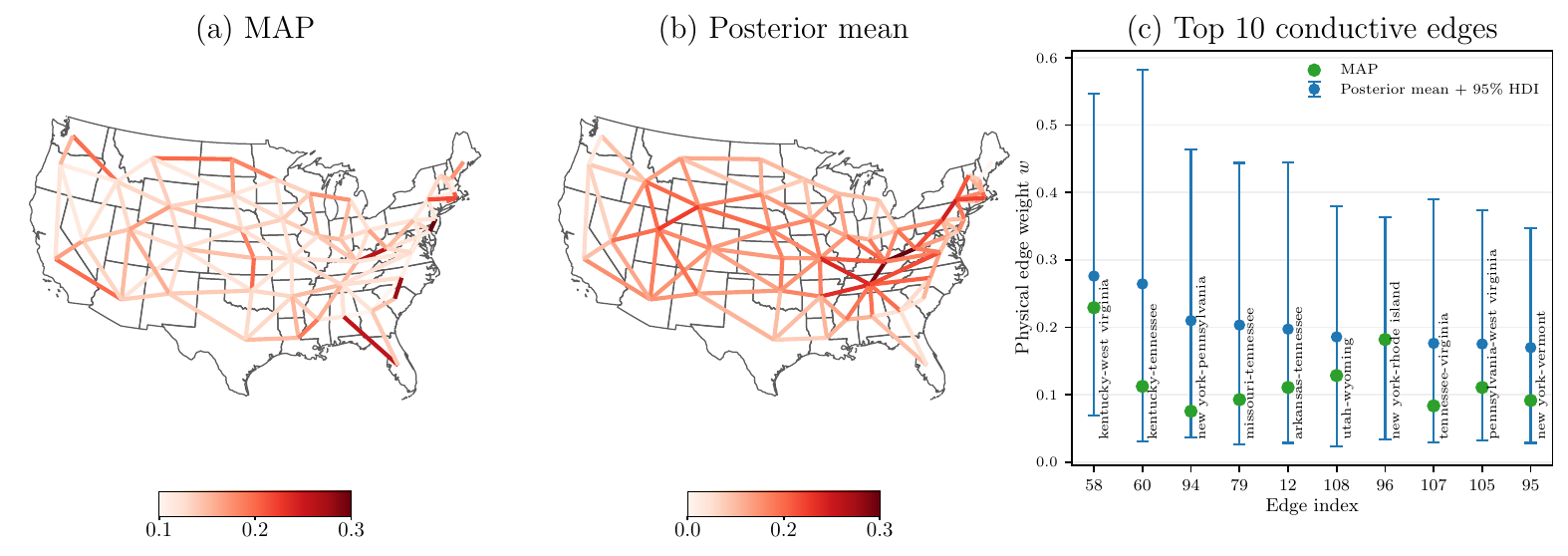}
    \caption{Graph-weight inference from the COVID-19 data using the
    non-stationary nonlinear reaction--diffusion model:
    (a) MAP estimate, (b) posterior mean, and
    (c) posterior edge-wise uncertainty.}
    \label{fig:heat_covid_nonlinearreaction}
\end{figure}

The inferred graph weights for the nonlinear reaction--diffusion model
are shown in \Cref{fig:heat_covid_nonlinearreaction}. We report the MAP
estimate, posterior mean, and the 10 most conductive edges together with
their $95\%$ HDIs. Compared with the synthetic examples, the MAP and
posterior mean exhibit more noticeable differences. The inferred weights are heterogeneous, with few strongly conductive connections emerging from a lower-conductivity background. 

Within the simplified graph reaction--diffusion model, the inferred
weights represent effective inter-state couplings that best explain the
observed spatiotemporal evolution of cumulative COVID-19 cases. Several
of the strongest inferred connections are incident to Kentucky and New
York. These connections should not be interpreted as direct measures of mobility or epidemiological transmission between states, but rather as effective couplings induced by the assumed dynamical model. Nevertheless, the results demonstrate that the framework infers heterogeneous graph connectivity and its uncertainty from real spatiotemporal observations.

This application has several limitations. The geographical graph
represents inter-state connectivity primarily through shared borders and
therefore neglects long-range pathways such as air travel. In addition,
the reaction parameters are fixed rather than inferred, and the
reaction--diffusion model provides a simplified representation of
epidemic spread. More realistic mobility networks and additional
parameter uncertainty would therefore be required for an epidemiological
interpretation of individual inferred connections.
\section{Conclusions}
We developed B-GRASP, a Bayesian framework for inverse problems involving
stochastic dynamics on weighted graphs, in which the graph edge weights
are treated as uncertain parameters that determine the graph differential
operator and hence the evolution of the nodal states. Motivated by the
connection between differential operators in PDEs and SPDEs and their
graph counterparts, B-GRASP enables the weighted graph underlying the
dynamics to be inferred from noisy observations rather than prescribed
a priori.

The framework uses a hierarchical formulation in which latent graph
variables determine positive edge weights and the associated graph
Laplacian, while latent Brownian variables represent realizations of the
stochastic forcing. Conditional on these variables, the stochastic graph dynamics reduce to a deterministic forward computation from which the likelihood and posterior distribution can be constructed. This formulation allows uncertainty associated with the unknown graph weights and stochastic forcing to be represented within a Bayesian model and applies to stationary and non-stationary dynamics as well as linear and nonlinear reaction–diffusion systems.
MAP estimation and NUTS sampling provide complementary point estimates
and posterior uncertainty quantification.

Numerical experiments of increasing complexity demonstrate the framework.
The one-dimensional inhomogeneous heat problem connects graph-weight
inference with classical coefficient estimation for continuous diffusion,
while synthetic stationary and non-stationary reaction--diffusion
problems illustrate inference under nonlinear dynamics and stochastic
forcing. The application to state-level COVID-19 data demonstrates the
framework with real spatio-temporal observations. Across the examples,
posterior distributions reveal uncertainty in inferred edge
weights not captured by point estimates alone and show that edge
identifiability depends on the extent to which individual connections
influence the observed dynamics.

Future work could explore improved scalability to larger graphs and longer time series, particularly when stochastic forcing introduces high-dimensional latent variables. Structured variational approximations and related methods
provide promising alternatives to full posterior sampling. More broadly,
B-GRASP provides a foundation for uncertainty-aware inference of dynamical
processes on weighted networks, with applications in transportation, energy
systems, epidemiology, biological networks, chip design, and materials and
drug discovery.

\section*{Acknowledgements}
In addition to the funding declared on the cover page, the authors thank Alexander Nikitin for his support with the COVID-19 data. The authors acknowledge the use of the large language models Claude Opus 5.5 and ChatGPT 5.6 to improve the clarity, grammar, and style of several sentences and to support data analysis and code development. The authors reviewed and edited all AI-assisted content as needed and take full responsibility for the content of this publication.

\section*{Author contributions}
Both authors made substantial intellectual contributions to the conception, design, and execution of the study and reviewed and approved the final version of the manuscript. \textbf{CRediT authorship contribution statement:} Christina Schenk and Babak Maboudi Afkham: Conceptualization, Methodology, Formal analysis, Investigation, Software, Data curation, Visualization, Writing – original draft, Writing – review \& editing, Funding acquisition.
\bibliographystyle{siamplain}
\bibliography{refs}

@book{Kuhl,
  author = {Kuhl, Ellen},
  year = {2021},
  title = {Computational Epidemiology},
  publisher = {Springer International Publishing Cham},
  isbn={978-3-030-82890-5}
}

@InProceedings{Nikitin2022,
  title = 	 { Non-separable Spatio-temporal Graph Kernels via {SPDEs} },
  author =       {Nikitin, Alexander V. and John, St and Solin, Arno and Kaski, Samuel},
  booktitle = 	 {Proceedings of The 25th International Conference on Artificial Intelligence and Statistics},
  pages = 	 {10640--10660},
  year = 	 {2022},
  editor = 	 {Camps-Valls, Gustau and Ruiz, Francisco J. R. and Valera, Isabel},
  volume = 	 {151},
  series = 	 {Proceedings of Machine Learning Research},
  month = 	 {28--30 Mar},
  publisher =    {PMLR},
  url = 	 {https://proceedings.mlr.press/v151/nikitin22a.html}
}

@InProceedings{Borovitskiy2021,
  title = 	 { Mat{é}rn {G}aussian Processes on Graphs },
  author =       {Borovitskiy, Viacheslav and Azangulov, Iskander and Terenin, Alexander and Mostowsky, Peter and Deisenroth, Marc and Durrande, Nicolas},
  booktitle = 	 {Proceedings of The 24th International Conference on Artificial Intelligence and Statistics},
  pages = 	 {2593--2601},
  year = 	 {2021},
  editor = 	 {Banerjee, Arindam and Fukumizu, Kenji},
  volume = 	 {130},
  series = 	 {Proceedings of Machine Learning Research},
  month = 	 {13--15 Apr},
  publisher =    {PMLR},
  url = 	 {https://proceedings.mlr.press/v130/borovitskiy21a.html}
}

@article{perri_bayesian_2023,
	title = {Bayesian inference of transition matrices from incomplete graph data with a topological prior},
	volume = {12},
	issn = {2193-1127},
	doi = {10.1140/epjds/s13688-023-00416-3},
	language = {eng},
	number = {1},
	journal = {EPJ data science},
	author = {Perri, Vincenzo and Petrović, Luka V. and Scholtes, Ingo},
	year = {2023},
	pmid = {37840552},
	pmcid = {PMC10567898},
	pages = {48},
}

@inproceedings{pmlr-v206-borovitskiy23arn,
	author = {Borovitskiy, Viacheslav and Karimi, Mohammad Reza and Somnath, Vignesh Ram and Krause, Andreas},
	booktitle = {Proceedings of The 26th International Conference on Artificial Intelligence and Statistics (AISTATS 2023)},
	month = {25--27 Apr},
	publisher = {PMLR},
	series = {Proceedings of Machine Learning Research},
	title = {Isotropic {G}aussian Processes on Finite Spaces of Graphs},
	volume = {206},
	year = {2023}}

@article{Harlim_2022,
doi = {10.1088/1361-6420/ac3994},
url = {https://dx.doi.org/10.1088/1361-6420/ac3994},
year = {2022},
month = {jan},
publisher = {IOP Publishing},
volume = {38},
number = {3},
pages = {035006},
author = {Harlim, John and Jiang, Shixiao W and Kim, Hwanwoo and Sanz-Alonso, Daniel},
title = {Graph-based prior and forward models for inverse problems on manifolds with boundaries},
journal = {Inverse Problems}
}

@book{Rasmussen2006Gaussian,
  author = {Rasmussen, Carl Edward and Williams, Christopher K. I.},
  publisher = {The MIT Press},
  title = {Gaussian Processes for Machine Learning},
  year = 2006
}

@ARTICLE{Belkin2003,
  author={Belkin, Mikhail and Niyogi, Partha},
  journal={Neural Computation}, 
  title={Laplacian Eigenmaps for Dimensionality Reduction and Data Representation}, 
  year={2003},
  volume={15},
  number={6},
  pages={1373-1396},
  doi={10.1162/089976603321780317}}

@ARTICLE{Yan2007,
  author={Yan, Shuicheng and Xu, Dong and Zhang, Benyu and Zhang, Hong-jiang and Yang, Qiang and Lin, Stephen},
  journal={IEEE Transactions on Pattern Analysis and Machine Intelligence}, 
  title={Graph Embedding and Extensions: A General Framework for Dimensionality Reduction}, 
  year={2007},
  volume={29},
  number={1},
  pages={40-51},
  doi={10.1109/TPAMI.2007.250598}}

@misc{sanzalonso2021spdeapproachmaternfields,
      title={The {SPDE} Approach to {Mat\'ern} Fields: Graph Representations}, 
      author={Daniel Sanz-Alonso and Ruiyi Yang},
      year={2021},
      eprint={2004.08000},
      archivePrefix={arXiv},
      primaryClass={stat.ME},
      url={https://arxiv.org/abs/2004.08000}, 
}

@article{Lindgren2011,
author = {Lindgren, Finn and Rue, Håvard and Lindström, Johan},
title = {An explicit link between {G}aussian fields and {G}aussian {M}arkov random fields: the stochastic partial differential equation approach},
journal = {Journal of the Royal Statistical Society: Series B (Statistical Methodology)},
volume = {73},
number = {4},
pages = {423-498},
doi = {https://doi.org/10.1111/j.1467-9868.2011.00777.x},
url = {https://rss.onlinelibrary.wiley.com/doi/abs/10.1111/j.1467-9868.2011.00777.x},
eprint = {https://rss.onlinelibrary.wiley.com/doi/pdf/10.1111/j.1467-9868.2011.00777.x},
year = {2011}
}

@article{Lindgren2015,
 title={Bayesian Spatial Modelling with {R-INLA}},
 volume={63},
 url={https://www.jstatsoft.org/index.php/jss/article/view/v063i19},
 doi={10.18637/jss.v063.i19},
 number={19},
 journal={Journal of Statistical Software},
 author={Lindgren, Finn and Rue, Håvard},
 year={2015},
 pages={1–25}
}

@InProceedings{Smola2003,
author="Smola, Alexander J.
and Kondor, Risi",
editor="Sch{\"o}lkopf, Bernhard
and Warmuth, Manfred K.",
title="Kernels and Regularization on Graphs",
booktitle="Learning Theory and Kernel Machines",
year="2003",
publisher="Springer Berlin Heidelberg",
address="Berlin, Heidelberg",
pages="144--158",
isbn="978-3-540-45167-9"
}

@book{solin_stochastic_2016,
	title = {Stochastic {Differential} {Equation} {Methods} for {Spatio}-{Temporal} {Gaussian} {Process} {Regression}},
	isbn = {978-952-60-6711-7},
	url = {https://aaltodoc.aalto.fi/handle/123456789/19842},
	language = {en},
	publisher = {Aalto University},
	author = {Solin, Arno},
	year = {2016},
	note = {ISSN: 1799-4942 (electronic)},
}

@article{whittle1963,
  author = {Peter Whittle},
  title = {Stochastic processes in several dimensions},
  journal = {Bulletin of the International Statistical Institute},
  volume = {40},
  pages = {974--994},
  year = {1963},
  url = {https://www.jstor.org/stable/10.2307/1402352}
}

@book{hooten2017animal,
  title={Animal Movement: Statistical Models for Telemetry Data},
  author={Hooten, M. B. and Johnson, D. S. and McClintock, B. T. and Morales, J. M.},
  year={2017},
  publisher={CRC Press},
  address={Boca Raton, Florida}
}

@book{burrough2015principles,
  title={Principles of Geographical Information Systems},
  author={Burrough, P. A. and McDonnell, R. A. and Lloyd, C. D.},
  year={2015},
  publisher={Oxford University Press},
  address={Oxford},
  edition={Second},
  isbn={0198742843}
}

@article{moraga2017geostatistical,
  title={A geostatistical model for combined analysis of point-level and area-level data using {INLA} and {SPDE}},
  author={Moraga, Paula and Cramb, Stella M. and Mengersen, Kerrie L. and Pagano, Marina},
  journal={Spatial Statistics},
  volume={21},
  pages={27--41},
  year={2017},
  publisher={Elsevier},
  doi={10.1016/j.spasta.2017.04.003}
}

@book{gilks1996markov,
  title={Markov {C}hain {M}onte {C}arlo in {P}ractice},
  author={Gilks, Walter R. and Richardson, Sylvia and Spiegelhalter, David J.},
  year={1996},
  publisher={Chapman and Hall/CRC},
  address={London},
  isbn={9780412055515}
}

@book{robert2013monte,
  title={Monte {C}arlo Statistical Methods},
  author={Robert, Christian P. and Casella, George},
  year={2013},
  edition={2nd},
  publisher={Springer},
  address={New York},
  isbn={9781441915756}
}

@book{neal2011mcmc,
	address = {New York},
	title = {Handbook of {Markov} {Chain} {Monte} {Carlo}},
	isbn = {978-0-429-13850-8},
	publisher = {Chapman and Hall/CRC},
	editor = {Brooks, Steve and Gelman, Andrew and Jones, Galin and Meng, Xiao-Li},
	month = may,
	year = {2011},
	doi = {10.1201/b10905},
}

@article{hoffman2014no,
  title={The {N}o-{U}-{T}urn {S}ampler: {A}daptively {S}etting {P}ath {L}engths in {H}amiltonian {M}onte {C}arlo},
  author={Hoffman, Matthew D. and Gelman, Andrew},
  journal={Journal of Machine Learning Research},
  volume={15},
  pages={1593--1623},
  year={2014}
}

@article{dunson2022graph,
  title={Graph Based {G}aussian Processes on Restricted Domains},
  author={Dunson, David B and Wu, Hau-Tieng and Wu, Nan},
  journal={Journal of the Royal Statistical Society: Series B (Statistical Methodology)},
  volume={84},
  number={2},
  pages={414--434},
  year={2022},
  url={https://academic.oup.com/jrsssb/article/84/2/414/7056155}
}

@inproceedings{goldberg1997regression,
  title={Regression with Input-dependent Noise: A {G}aussian Process Treatment},
  author={Goldberg, Paul W and Williams, Christopher K I and Bishop, Christopher M},
  booktitle={Advances in Neural Information Processing Systems},
  year={1997},
  url={http://www.cs.ox.ac.uk/people/paul.goldberg/papers/gwb-nips97.pdf}
}

@inproceedings{blanco2021evolving,
  title={Evolving-Graph {G}aussian Processes},
  author={Blanco-Mulero, David and Heinonen, Markus and Kyrki, Ville},
  booktitle={Time Series Workshop at ICML},
  year={2021},
  url={https://roseyu.com/time-series-workshop/submissions/2021/TSW-ICML2021_paper_21.pdf}
}

@incollection{rasmussen2004gaussian,
	address = {Berlin, Heidelberg},
	title = {Gaussian {Processes} in {Machine} {Learning}},
	isbn = {978-3-540-28650-9},
	url = {https://doi.org/10.1007/978-3-540-28650-9_4},
	language = {en},
	booktitle = {Advanced {Lectures} on {Machine} {Learning}: {ML} {Summer} {Schools} 2003, {Canberra}, {Australia}, {February} 2 - 14, 2003, {Tübingen}, {Germany}, {August} 4 - 16, 2003, {Revised} {Lectures}},
	publisher = {Springer},
	author = {Rasmussen, Carl Edward},
	editor = {Bousquet, Olivier and von Luxburg, Ulrike and Rätsch, Gunnar},
	year = {2004},
	doi = {10.1007/978-3-540-28650-9_4},
	pages = {63--71},
}

@article{paszke2019pytorch,
  author    = {Adam Paszke and Sam Gross and Francisco Massa and Adam Lerer and James Bradbury
               and Gregory Chanan and Trevor Killeen and Zeming Lin and Natalia Gimelshein and
               Luca Antiga and Alban Desmaison and Andreas K{\"o}pf and Edward Yang and Zachary DeVito
               and Martin Raison and Alykhan Tejani and Sasank Chilamkurthy and Benoit Steiner
               and Lu Fang and Junjie Bai and Soumith Chintala},
  title     = {PyTorch: An Imperative Style, High-Performance Deep Learning Library},
  journal   = {Advances in Neural Information Processing Systems},
  volume    = {32},
  year      = {2019},
  pages     = {8024--8035}
}

@book{chung1997spectral,
  title={Spectral graph theory},
  author={Chung, Fan RK},
  volume={92},
  year={1997},
  publisher={American Mathematical Soc.}
}

@book{lions1971optimal,
  title={Optimal control of systems governed by partial differential equations},
  author={Lions, Jacques Louis},
  volume={170},
  year={1971},
  publisher={Springer}
}

@book{ciarlet2025linear,
  title={Linear and nonlinear functional analysis with applications},
  author={Ciarlet, Philippe G},
  year={2025},
  publisher={SIAM}
}

@article{liu1989limited,
  title={On the limited memory {BFGS} method for large scale optimization},
  author={Liu, Dong C and Nocedal, Jorge},
  journal={Mathematical Programming},
  volume={45},
  number={1-3},
  pages={503--528},
  year={1989},
  publisher={Springer}
}

@article{bingham2019pyro,
  title={Pyro: Deep Universal Probabilistic Programming},
  author={Bingham, Eli and Chen, Jonathan P. and Jankowiak, Martin and Obermeyer, Fritz and Pradhan, Neeraj and Karaletsos, Theofanis and Singh, Rohit and Szerlip, Paul A. and Horsfall, Paul and Goodman, Noah D.},
  journal={Journal of Machine Learning Research},
  volume={20},
  number={28},
  pages={1--6},
  year={2019}
}

@misc{NikitinCovid,
    year={2023},
    author = {Alexander Nikitin},
    title = {covid19-on-graphs},
    url={https://github.com/AlexanderVNikitin/covid19-on-graphs}
    }

@misc{NYtimesCoviddata,
    year={2023},
    author = {S. Almukhtar et al.},
    title = {Coronavirus ({C}ovid-19) Data in the {U}nited {S}tates},
    url={https://github.com/nytimes/covid-19-data}
    }

@article{Schenk2024,
author = {Schenk, Christina and Vasudevan, Aditya and Haranczyk, Maciej and Romero, Ignacio},
title = {Model-based reinforcement learning control of reaction-diffusion problems},
journal = {Optimal Control Applications and Methods},
volume = {45},
number = {6},
pages = {2897-2914},
doi = {https://doi.org/10.1002/oca.3196},
url = {https://onlinelibrary.wiley.com/doi/abs/10.1002/oca.3196},
eprint = {https://onlinelibrary.wiley.com/doi/pdf/10.1002/oca.3196},
year = {2024}
}

@book{kaipio2005statistical,
  title={Statistical and computational inverse problems},
  author={Kaipio, Jari P and Somersalo, Erkki},
  year={2005},
  publisher={Springer}
}

@article{bolin_gaussian_2024,
	title = {Gaussian {Whittle}–{Matérn} fields on metric graphs},
	volume = {30},
	issn = {1350-7265},
	url = {https://projecteuclid.org/journals/bernoulli/volume-30/issue-2/Gaussian-WhittleMat%c3%a9rn-fields-on-metric-graphs/10.3150/23-BEJ1647.full},
	doi = {10.3150/23-BEJ1647},
	number = {2},
	journal = {Bernoulli},
	publisher = {Bernoulli Society for Mathematical Statistics and Probability},
	author = {Bolin, David and Simas, Alexandre B. and Wallin, Jonas},
	month = may,
	year = {2024},
	pages = {1611--1639},
}

@article{bolin_statistical_2026,
	title = {Statistical inference for {Gaussian} {Whittle}–{Matérn} fields on metric graphs},
	issn = {1369-7412},
	url = {https://doi.org/10.1093/jrsssb/qkag074},
	doi = {10.1093/jrsssb/qkag074},
	journal = {Journal of the Royal Statistical Society Series B: Statistical Methodology},
	author = {Bolin, David and Simas, Alexandre B and Wallin, Jonas},
	month = may,
	year = {2026},
	pages = {qkag074},
}

@article{berild_non-stationary_2024,
	title = {Non-stationary spatio-temporal modeling using the stochastic advection–diffusion equation},
	volume = {64},
	issn = {2211-6753},
	url = {https://www.sciencedirect.com/science/article/pii/S2211675324000587},
	doi = {10.1016/j.spasta.2024.100867},
	journal = {Spatial Statistics},
	author = {Berild, Martin Outzen and Fuglstad, Geir-Arne},
	month = dec,
	year = {2024},
	pages = {100867},
}

@article{clarotto_spde_2024,
	title = {The {SPDE} approach for spatio-temporal datasets with advection and diffusion},
	volume = {62},
	issn = {2211-6753},
	url = {https://www.sciencedirect.com/science/article/pii/S2211675324000381},
	doi = {10.1016/j.spasta.2024.100847},
	journal = {Spatial Statistics},
	author = {Clarotto, Lucia and Allard, Denis and Romary, Thomas and Desassis, Nicolas},
	month = aug,
	year = {2024},
	pages = {100847},
}

@article{mostowsky_geometrickernels_2025,
	title = {The {GeometricKernels} {Package}: {Heat} and {Matérn} {Kernels} for {Geometric} {Learning} on {Manifolds}, {Meshes}, and {Graphs}},
	volume = {26},
	issn = {1533-7928},
	shorttitle = {The {GeometricKernels} {Package}},
	url = {http://jmlr.org/papers/v26/24-1185.html},
	number = {276},
	journal = {Journal of Machine Learning Research},
	author = {Mostowsky, Peter and Dutordoir, Vincent and Azangulov, Iskander and Jaquier, Noémie and Hutchinson, Michael John and Ravuri, Aditya and Rozo, Leonel and Terenin, Alexander and Borovitskiy, Viacheslav},
	year = {2025},
	pages = {1--14},
}

@inproceedings{alain_gaussian_2024,
	title = {Gaussian {Processes} on {Cellular} {Complexes}},
	issn = {2640-3498},
	url = {https://proceedings.mlr.press/v235/alain24a.html},
	language = {en},
	booktitle = {Proceedings of the 41st {International} {Conference} on {Machine} {Learning}},
	publisher = {PMLR},
	author = {Alain, Mathieu and Takao, So and Paige, Brooks and Deisenroth, Marc Peter},
	month = jul,
	year = {2024},
	pages = {879--905},
}

@article{xumarkovich_uncertainty_nodate,
	title = {Uncertainty {Estimation} on {Graphs} with {Structure} {Informed} {Stochastic} {Partial} {Differential} {Equations}},
	language = {en},
	author = {Xu, Fred and Markovich, Thomas},
}

@article{SchenkPortilloRomero2023,
author = {Schenk, Christina and Portillo, David and Romero, Ignacio},
title = {Linking discrete and continuum diffusion models: Well-posedness and stable finite element discretizations},
journal = {International Journal for Numerical Methods in Engineering},
volume = {124},
number = {9},
pages = {2105-2121},
doi = {https://doi.org/10.1002/nme.7204},
url = {https://onlinelibrary.wiley.com/doi/abs/10.1002/nme.7204},
eprint = {https://onlinelibrary.wiley.com/doi/pdf/10.1002/nme.7204},
year = {2023}
}

@article{stuart2010inverse,
  title = {Inverse problems: A {B}ayesian perspective},
  author = {Stuart, Andrew M.},
  journal = {Acta Numerica},
  volume = {19},
  pages = {451--559},
  year = {2010},
  publisher = {Cambridge University Press},
  doi = {10.1017/S0962492910000061},
  url = {https://www.cambridge.org/core/journals/acta-numerica/article/inverse-problems-a-bayesian-perspective/587A3A0D480A1A7C2B1B284BCEDF7E23}
}

@Article{schenk2026ac,
  author       = {Schenk, C. and Romero, I.},
  title        = {A Framework for the {B}ayesian Calibration of Complex and Data-Scarce Models in Applied Sciences},
  journal      = {Arch Computat Methods Eng},
  year         = {2026},
  doi          = {https://doi.org/10.1007/s11831-026-10665-w}
}

@article{lindgren2011explicit,
  title={An explicit link between Gaussian fields and Gaussian Markov random fields: the stochastic partial differential equation approach},
  author={Lindgren, Finn and Rue, H{\aa}vard and Lindstr{\"o}m, Johan},
  journal={Journal of the royal statistical society: Series b (statistical methodology)},
  volume={73},
  number={4},
  pages={423--498},
  year={2011},
  publisher={Wiley Online Library}
}

@article{zhi_gaussian_2023,
	title = {Gaussian {Processes} on {Graphs} {Via} {Spectral} {Kernel} {Learning}},
	volume = {9},
	copyright = {https://ieeexplore.ieee.org/Xplorehelp/downloads/license-information/IEEE.html},
	issn = {2373-776X, 2373-7778},
	url = {https://ieeexplore.ieee.org/document/10093993/},
	doi = {10.1109/TSIPN.2023.3265160},
	language = {en},
	journal = {IEEE Transactions on Signal and Information Processing over Networks},
	author = {Zhi, Yin-Cong and Ng, Yin Cheng and Dong, Xiaowen},
	year = {2023},
	pages = {304--314},
}

@article{hein2007graph,
  title={Graph {L}aplacians and their convergence on random neighborhood graphs.},
  author={Hein, Matthias and Audibert, Jean-Yves and Luxburg, Ulrike von},
  journal={Journal of Machine Learning Research},
  volume={8},
  number={6},
  year={2007}
}

@inproceedings{opolka_adaptive_2022,
	title = {Adaptive {Gaussian} {Processes} on {Graphs} via {Spectral} {Graph} {Wavelets}},
	issn = {2640-3498},
	url = {https://proceedings.mlr.press/v151/opolka22a.html},
	language = {en},
	booktitle = {Proceedings of {The} 25th {International} {Conference} on {Artificial} {Intelligence} and {Statistics}},
	publisher = {PMLR},
	author = {Opolka, Felix and Zhi, Yin-Cong and Lió, Pietro and Dong, Xiaowen},
	month = may,
	year = {2022},
	pages = {4818--4834},
}

@misc{maboudiafkham2026bgrasp,
  author       = {Maboudi Afkham, Babak and Schenk, Christina},
  title        = {{B-GRASP}: Bayesian GRAph Inference with SPDE priors},
  year         = {2026},
  howpublished = {\url{https://github.com/babakmaboudi/B-GRASP}},
  note         = {GitHub repository}
}
\newpage
\section*{\large{Supplementary Materials:  B-GRASP: A Bayesian Framework for Inferring\\Graph Weights from SPDE-Inspired Dynamics}}
\subsection{Derivation for the linearized reaction term}
\subsubsection{Linearized reaction term}
In addition to the nonlinear reaction term, we consider its linearization around a reference state $F_0$. A first-order Taylor expansion gives
\begin{equation}
	\mathcal{R}(F)
	\approx
	\mathcal{R}(F_0)\boldsymbol{1}
	+
	\mathcal{R}'(F_0)(F-F_0\boldsymbol{1}),
\end{equation}
which can equivalently be written in affine form as
\begin{equation}
	\label{eq:linearized-reaction}
	\begin{aligned}
		\mathcal{R}(F)
		&\approx
		c_{\mathrm{const}}\boldsymbol{1}
		+
		\mathcal{R}'(F_0)F,
		\\
		\text{where}\qquad
		\mathcal{R}'(F_0)
		&=
		(\psi-\gamma)-2\psi F_0,
		\\
		c_{\mathrm{const}}
		&=
		\mathcal{R}(F_0)
		-
		\mathcal{R}'(F_0)F_0.
	\end{aligned}
\end{equation}
Here, $\boldsymbol{1}\in\mathbb{R}^{N_{\mathcal V}}$ denotes the vector of ones.
For the quadratic SIS reaction term, the constant term simplifies to
\begin{equation} \label{eq:c_const}
	c_{\mathrm{const}}
	=
	\psi F_0^2.
\end{equation}

We choose the reference state $F_0$ to be the endemic equilibrium, which exists for $\psi>\gamma$. Since $\mathcal{R}(F_0)=0$, the derivative of the reaction term at this equilibrium simplifies to
\begin{equation}
	\label{eq:reaction-derivative-equilibrium}
	\mathcal{R}'(F_0)
	=
	-(\psi-\gamma).
\end{equation}
The negative derivative reflects the local stability of the endemic equilibrium with respect to the reaction dynamics.
\subsubsection{Discretization of the linearized model}
The linearized reaction model leads to a linear stochastic recurrence, which permits both explicit and implicit time discretizations. For the linearized reaction term in (SM1.2), the drift takes the form

For the linearized reaction model in \eqref{eq:linearized-reaction}, the drift becomes
\begin{equation}
	-\boldsymbol{L}_{\kappa,\alpha}(\boldsymbol{w})F
	+\beta
	\mathcal{R}'(F_0)F
	+
	c_{\mathrm{const}}\boldsymbol{1},
\end{equation}
where $c_{\mathrm{const}}$ is defined \eqref{eq:c_const}. We therefore define the effective linear operator
\begin{equation}
	\label{eq:effective-operator}
	\boldsymbol{L}_{\kappa,\alpha}^{\mathrm{eff}}(\boldsymbol{w})
	=
	\boldsymbol{L}_{\kappa,\alpha}(\boldsymbol{w})
	-
	\beta\mathcal{R}'(F_0)\boldsymbol{I}_{N_{\mathcal V}},
\end{equation}
such that the linearized drift can be written as
\begin{equation}
	-\boldsymbol{L}_{\kappa,\alpha}^{\mathrm{eff}}(\boldsymbol{w})F
	+
	\beta c_{\mathrm{const}}\boldsymbol{1}.
\end{equation}

An explicit Euler--Maruyama discretization of the linearized process is given by
\begin{equation}
	\label{eq:linear-explicit}
	F(t_i;\boldsymbol{w})
	=
	F(t_{i-1};\boldsymbol{w})
	-
	\Delta t\,
	\boldsymbol{L}_{\kappa,\alpha}^{\mathrm{eff}}(\boldsymbol{w})
	F(t_{i-1};\boldsymbol{w})
	+
	\Delta t\,
	\beta c_{\mathrm{const}}\boldsymbol{1}
	+
	\sqrt{\Delta t}\,
	\boldsymbol{\Gamma}\boldsymbol{\varepsilon}_{i-1}.
\end{equation}
Alternatively, evaluating the linear drift implicitly at $t_i$ gives
\begin{equation}
	\label{eq:linear-implicit}
	F(t_i;\boldsymbol{w})
	=
	F(t_{i-1};\boldsymbol{w})
	-
	\Delta t\,
	\boldsymbol{L}_{\kappa,\alpha}^{\mathrm{eff}}(\boldsymbol{w})
	F(t_i;\boldsymbol{w})
	+
	\Delta t\,
	\beta c_{\mathrm{const}}\boldsymbol{1}
	+
	\sqrt{\Delta t}\,
	\boldsymbol{\Gamma}\boldsymbol{\varepsilon}_{i-1}.
\end{equation}
Equivalently, $F(t_i;\boldsymbol{w})$ is obtained by solving the linear system
\begin{equation}
	\label{eq:linear-implicit-system}
	\left(
	\boldsymbol{I}_{N_{\mathcal V}}
	+
	\Delta t\,
	\boldsymbol{L}_{\kappa,\alpha}^{\mathrm{eff}}(\boldsymbol{w})
	\right)
	F(t_i;\boldsymbol{w})
	=
	F(t_{i-1};\boldsymbol{w})
	+
	\Delta t\,\beta
	c_{\mathrm{const}}\boldsymbol{1}
	+
	\sqrt{\Delta t}\,
	\boldsymbol{\Gamma}\boldsymbol{\varepsilon}_{i-1}.
\end{equation}
\subsection{Additional results and details for the 1D inverse problem}
\subsubsection{1D heat problem and graph--finite-difference correspondence} \label{sec:planar-heat}
Let $\mathcal{G}$ be a path graph with $N_{\mathcal V}$ vertices (i.e., a tree consisting of a single simple path). We assign unit weight to all edges, i.e., $\boldsymbol{w}=\boldsymbol{1}$, where $\boldsymbol{1}\in\mathbb{R}^{N_{\mathcal E}}$ denotes the vector of ones and $N_{\mathcal E}$ is the number of edges. The corresponding adjacency matrix $\boldsymbol{A}$, degree matrix $\boldsymbol{D}$, and diagonal scaling matrix $\boldsymbol{S}:=\boldsymbol{D}^{-1/2}$ are then constructed as described in the preceding sections.
\begin{equation}
	\boldsymbol{A}(\boldsymbol{1}) =
	\begin{pmatrix}
		0 & 1 & 0 & \cdots & 0 \\
		1 & 0 & 1 & \ddots & \vdots \\
		0 & 1 & 0 & \ddots & 0 \\
		\vdots & \ddots & \ddots & \ddots & 1 \\
		0 & \cdots & 0 & 1 & 0
	\end{pmatrix},
	\boldsymbol{D}(\boldsymbol{1}) =
	\operatorname{diag}(1,2,\ldots,2,1),
	\boldsymbol{S}(\boldsymbol{1}) =
	\operatorname{diag}\left(1,\frac{1}{\sqrt{2}},\ldots,\frac{1}{\sqrt{2}},1\right),
\end{equation}
where diag$(\cdot)$ creates a diagonal matrix from a vector. The graph Laplacian $\boldsymbol{L}(\boldsymbol{w})$ and the symmetric normalized graph Laplacian $\widetilde{\boldsymbol{L}}(\boldsymbol{w})$, become, for $\boldsymbol{w}=\boldsymbol{1}$,
\begin{equation}
	\boldsymbol{L}(\boldsymbol{1}) =
	\begin{pmatrix}
		1 & -1 & 0 & \cdots & 0 \\
		-1 & 2 & -1 & \ddots & \vdots \\
		0 & -1 & 2 & \ddots & 0 \\
		\vdots & \ddots & \ddots & \ddots & -1 \\
		0 & \cdots & 0 & -1 & 1
	\end{pmatrix},
	\;
	\widetilde{\boldsymbol{L}}(\boldsymbol{1}) =
	\begin{pmatrix}
		1 & -1/\sqrt{2} & 0 & \cdots & 0 \\
		-1/\sqrt{2} & 1 & -1/2 & \ddots & \vdots \\
		0 & -1/2 & 1 & \ddots & 0 \\
		\vdots & \ddots & \ddots & \ddots & -1/\sqrt{2} \\
		0 & \cdots & 0 & -1/\sqrt{2} & 1
	\end{pmatrix}.
\end{equation}
The operator $\boldsymbol{L}$ corresponds to the standard second-order centered FD discretization of $-\Delta$ on a uniform grid with spacing $\Delta x=1$, with a boundary treatment analogous to homogeneous Neumann (``free'') boundary conditions. In the graph setting, no values are prescribed at the endpoints and no edges, and hence no flux, extend beyond the boundary of the graph. The normalized operator $\widetilde{\boldsymbol{L}}$, however, does not correspond to the same FD discretization of $-\Delta$ with standard Dirichlet, Neumann, or Robin boundary conditions. Rather, the normalization introduces a degree-dependent scaling, which differs at the boundary because the endpoint vertices have a lower degree than the interior vertices. This normalization can alternatively be interpreted in terms of a degree-weighted vertex measure and leads to a boundary-modified stencil; see, e.g., \cite{chung1997spectral}.

For the unit-weight case $w_{i+1/2}\equiv 1$ and the choice $\Delta x=1$, the interior stencil becomes $2F_i-F_{i-1}-F_{i+1}$, and the FD operator $\boldsymbol{L}_w$ corresponds to the unnormalized graph Laplacian $\boldsymbol{L}(\boldsymbol{1})$. The main distinction lies in the treatment and interpretation of the boundary: for the graph Laplacian, the boundary behavior is encoded through the graph connectivity and the reduced degree of the endpoint vertices, whereas in the PDE discretization the homogeneous Neumann boundary conditions are imposed explicitly through zero-flux constraints.

We impose homogeneous Neumann boundary conditions at both endpoints. The boundary rows of $\boldsymbol{L}_w$ are obtained by imposing zero flux in $x=0$ and $x=L$, i.e., $w_{1/2}(F_1-F_0)=0$ and $w_{N-1/2}(F_N-F_{N-1})=0$, respectively.

The relationship between the normalized graph Laplacian and a divergence-form FD discretization is more subtle. In contrast to the unnormalized graph Laplacian, the degree normalization introduces a spatially dependent scaling through the vertex degrees and therefore does not directly correspond to the standard discretization of $-\frac{d}{dx}\left(w(x)\frac{df}{dx}\right)$. At the continuum level, this behavior can be related to diffusion operators defined with respect to a spatially varying reference measure. A representative form is
\begin{equation}
	\widetilde{\mathcal{L}}_{w}f
	:=
	-\frac{1}{\rho(x)}
	\frac{d}{dx}
	\left(
	w(x)\frac{df}{dx}
	\right),
\end{equation}
where $\rho(x)$ represents a spatially varying density or reference measure, analogous to the role played by the vertex degrees in the discrete setting. The precise correspondence depends on the choice of graph normalization and its continuum interpretation, and a detailed analysis of this relationship is beyond the scope of the present work \cite{hein2007graph}.

\subsubsection{Computational and posterior sampling details}
The Brownian variables $\boldsymbol{z}$ enter the time-discretized model through the stochastic forcing term, whereas the latent graph variables $\boldsymbol{x}$ enter through the transformation $\boldsymbol{w}=\mathcal{T}(\boldsymbol{x})$ and subsequently through the graph Laplacian $\boldsymbol{L}(\boldsymbol{w})$. The dependence of the computational forward map $\mathcal{F}$ on $\boldsymbol{x}$ is therefore nonlinear and involves differentiation through both the graph construction and the time-stepping procedure. Such parameter-to-solution derivatives are well studied in the context of PDE-constrained optimization and can be formulated in terms of G\^{a}teaux or Fr\'echet derivatives; see, e.g., \cite{lions1971optimal,ciarlet2025linear}. In the present work, rather than deriving these sensitivities explicitly, we compute the required derivatives $\nabla_{\boldsymbol{x}}\mathcal{F}$ and $\nabla_{\boldsymbol{z}}\mathcal{F}$ using the automatic differentiation functionality provided by PyTorch \cite{paszke2019pytorch}.

We compute the MAP estimate by maximizing the log-posterior. This optimization problem is solved using the limited-memory Broyden--Fletcher--Goldfarb--Shanno (L-BFGS) algorithm \cite{liu1989limited}, as implemented in PyTorch, with zero vectors used as the initial guess. Our sampling strategy comprises a warm-up phase of 1000 iterations, during which the step size is adapted, followed by a sampling phase in which the step size is fixed and $N_{\text{sample}}=2000$ samples are drawn to approximate the posterior distribution. In this work, the Pyro implementation of NUTS \cite{bingham2019pyro} is utilized. We then estimate ergodic averages.

\subsubsection{Additional posterior diagnostics}
We present trace plots of the NUTS samples for a subset of the graph weights in \Cref{fig:graphical_heat_trace}. The trace plots show no obvious signs of poor mixing for the selected edge weights, and similar behavior is observed for the remaining weights.
\begin{figure}[htb]
	\centering
	\includegraphics[width=.65\linewidth]{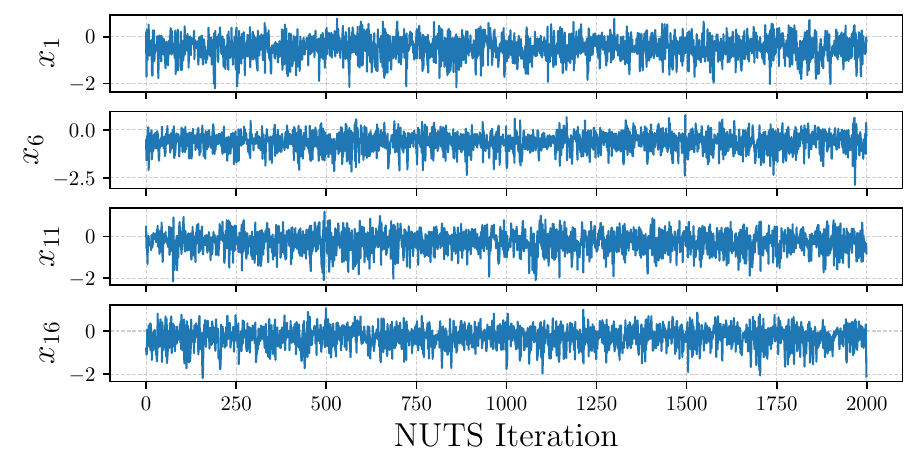}
	\caption{Trace plot for the NUTS samples of selected edge weights.}
	\label{fig:graphical_heat_trace}
\end{figure}
\subsubsection{Reconstruction of the stochastic forcing}
Although the primary objective of the inverse problem is to infer the edge weights, the proposed method also estimates the Brownian variables $\boldsymbol{z}$ that determine the discrete stochastic forcing in the heat equation. Since $\boldsymbol{z}$ enters the forward model linearly and is assigned a Gaussian prior, it could alternatively be marginalized from the posterior. Here, however, we retain $\boldsymbol{z}$ as an explicit latent variable, allowing the realization of the stochastic forcing to be estimated jointly with the unknown graph weights.

Figure \Cref{fig:graphical_heat_noise_estimation} compares the true Brownian variables with their posterior estimates. A color map is used to visualize $\boldsymbol{z}$, where each color represents the relative magnitude of the stochastic forcing at a particular vertex and time step. The ground-truth realization is shown in \Cref{fig:graphical_heat_noise_estimation}a, while the posterior mean is shown in \Cref{fig:graphical_heat_noise_estimation}b. Although the reconstruction is coarse, the posterior mean captures the larger-scale features of the ground-truth realization. This indicates that the stochastic forcing can, to some extent, be distinguished from the other sources of uncertainty in the inverse problem, including uncertainty in the graph weights and measurement noise.

\begin{figure}[htb]
	\centering
	\includegraphics[width=1\linewidth]{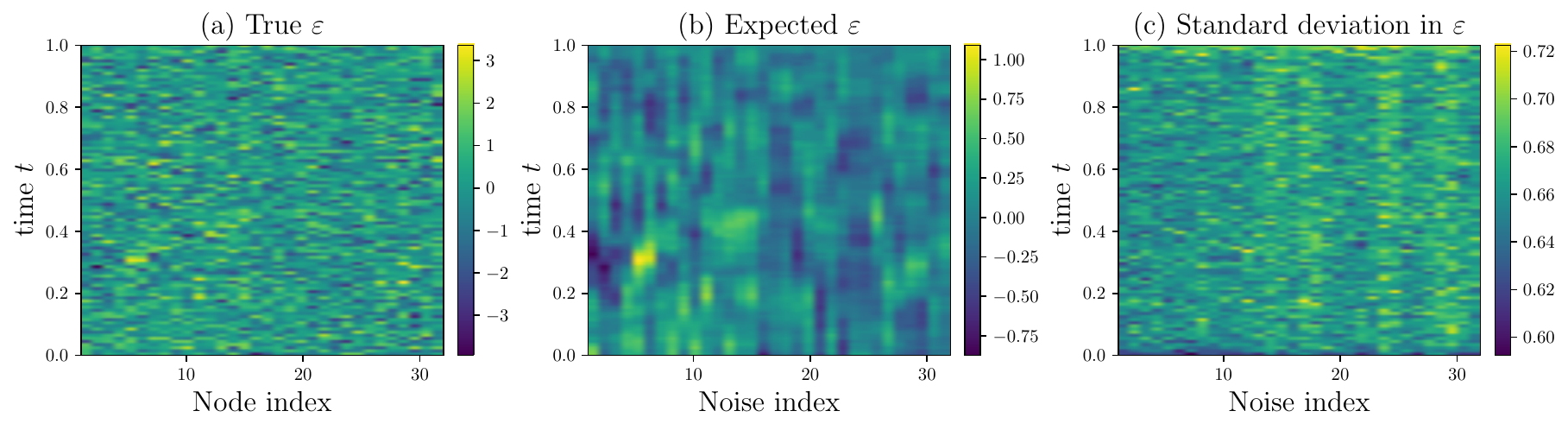}
	\caption{Estimation of the stochastic noise in the graph heat equation.}
	\label{fig:graphical_heat_noise_estimation}
\end{figure}

We observe that the MAP estimate of $\boldsymbol{z}$ closely matches the posterior mean. The proposed method also quantifies the uncertainty associated with the estimation of $\boldsymbol{z}$ through the pointwise posterior standard deviation, shown in \Cref{fig:graphical_heat_noise_estimation}c. The uncertainty generally increases over time, which may be attributed to later stochastic increments influencing a shorter portion of the observed trajectory and consequently being less constrained by the available observations.

\subsection{Additional results for the synthetic stationary graph
	reaction--diffusion problem with linear reaction}
\label{sec:sm-stationary-linear}

We repeat the synthetic stationary experiment described in
the main paper using the linear reaction model, with reaction weight $\beta=1$. The graph,
ground-truth edge weights, localized source, observation model, and
inference procedure are otherwise identical to those of the nonlinear
stationary experiment in the main text. The linear model provides a
local approximation of the nonlinear reaction dynamics considered
there.

\begin{figure}[htb]
	\centering
	\includegraphics[width=1\linewidth]{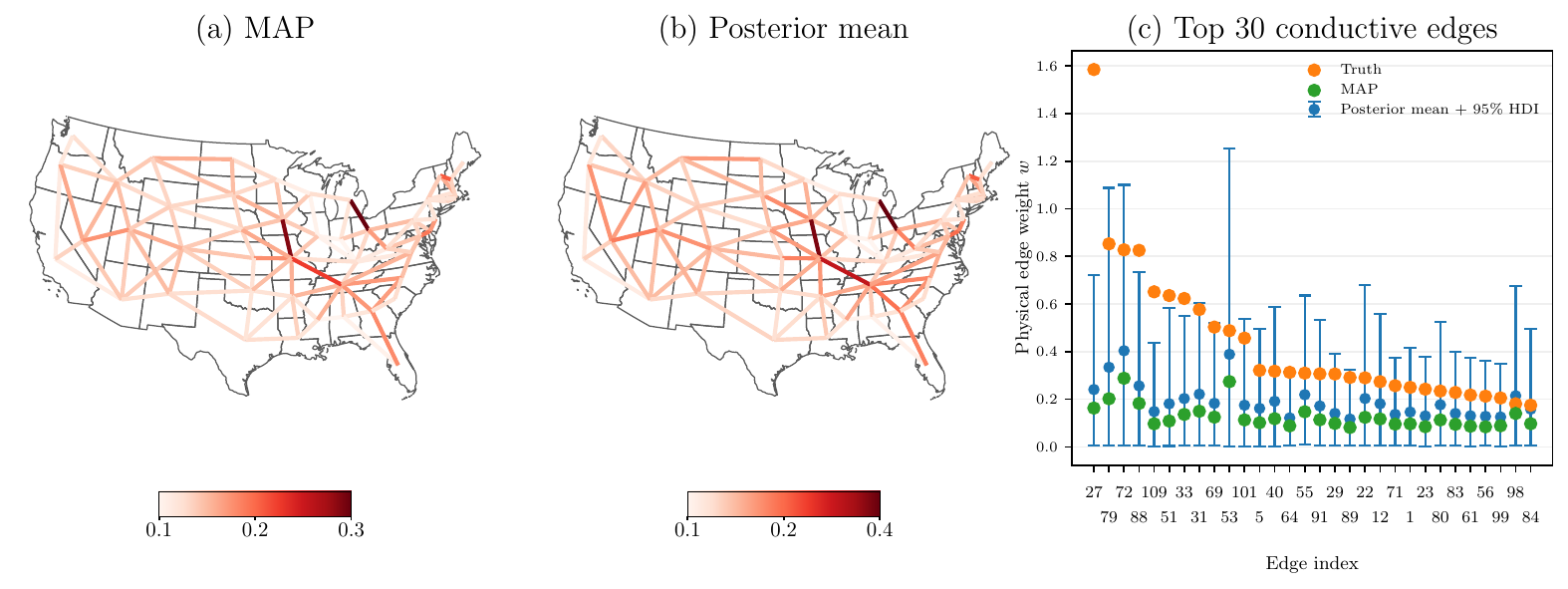}
	\caption{Graph-weight inference for the synthetic stationary
		reaction--diffusion problem with linear reaction dynamics:
		(a) MAP estimate, (b) posterior mean, and
		(c) posterior edge-wise uncertainty.}
	\label{fig:sm-stationary-linearreaction}
\end{figure}

\Cref{fig:sm-stationary-linearreaction} shows the MAP estimate,
posterior mean, and the 30 most conductive edges together with their
$95\%$ HDIs. As in the nonlinear stationary experiment, edge
identifiability depends strongly on graph excitation. As in the nonlinear stationary experiment, edge identifiability is governed primarily by graph excitation rather than conductivity alone. Low-conductivity background edges are generally recovered with narrow HDIs, whereas highly conductive edges are well identified only when they are sufficiently excited by the localized source. 
Some highly conductive edges farther from the source, such as edge 27,
remain weakly informed by the observations. Their conductivities have
comparatively little influence on the observed stationary state because
these parts of the graph are only weakly excited. Thus, as in the nonlinear case, temporal information improves inference for dynamically active edges but does not overcome the identifiability limitations associated with weak excitation. 

Compared with the linear model, the nonlinear stationary model reported
in the main text provides a sharper reconstruction of several highly
conductive edges near the localized source.

\subsection{Additional results for the synthetic non-stationary graph reaction/diffusion with linear reaction in the United States}
\label{sec:sm-nonstationary-linear}

We repeat the synthetic non-stationary experiment described in
the main manuscript using the linear reaction model with
$\beta=1$. All other aspects of the problem, including the graph
structure, ground-truth edge weights, initial condition, temporal
discretization, stochastic forcing, observation noise, and inference
procedure, are unchanged. In particular, the Brownian variables
$\boldsymbol{z}$ are inferred jointly with the latent graph variables
$\boldsymbol{x}$.

The initial condition, noise-free trajectory, and ground-truth graph
weights for the linear experiment are shown in
\Cref{fig:sm-heat-signal-linear}.
\begin{figure}[htb]
	\centering
	\includegraphics[width=1\linewidth]
	{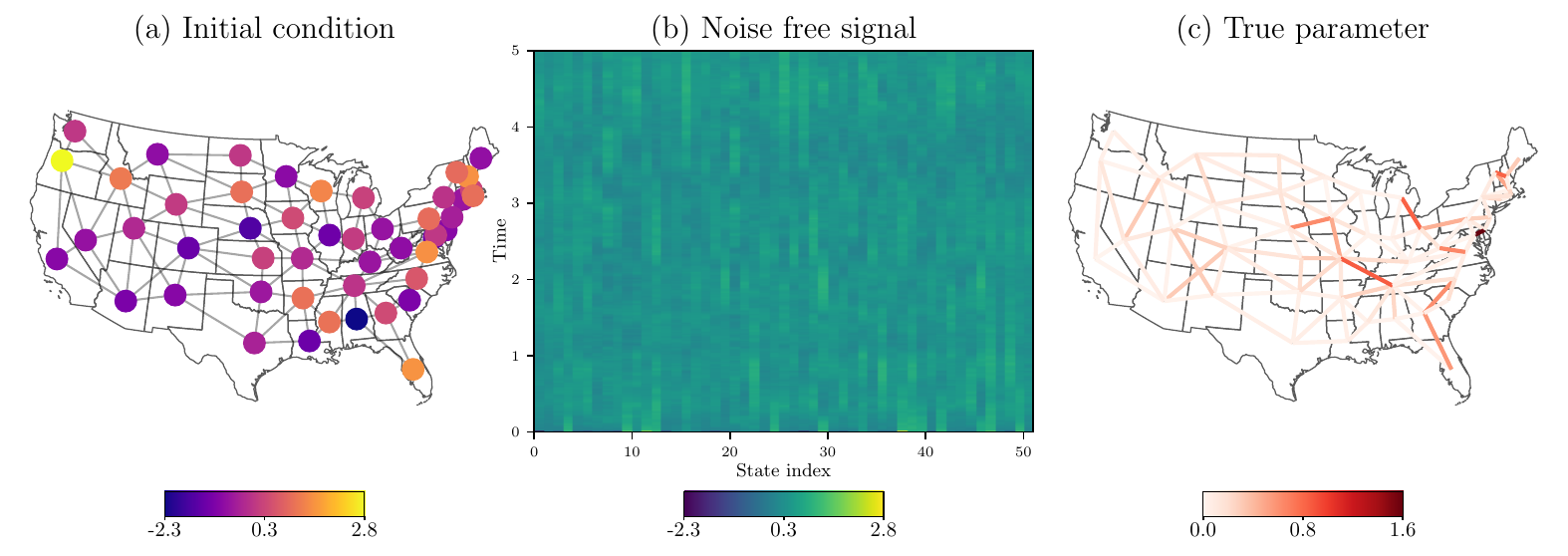}
	\caption{Synthetic non-stationary reaction--diffusion problem with
		linear reaction dynamics: (a) initial condition, (b) noise-free
		nodal trajectories, and (c) ground-truth graph edge weights.}
	\label{fig:sm-heat-signal-linear}
\end{figure}

\Cref{fig:sm-heat-linearreaction} shows the MAP estimate, posterior mean,
and the most conductive edges together with their $95\%$ HDIs. Many of
the dominant conductive pathways are identified by both the MAP and
posterior mean estimates. As in the nonlinear experiment reported in the
main text, the HDIs are generally narrower than in the corresponding
stationary problem, indicating improved identifiability when time-resolved
observations are available.

Some highly conductive edges nevertheless remain outside their
corresponding $95\%$ HDIs. Consistent with the stationary results, these
edges are only weakly excited by the dynamics and therefore have limited
influence on the observations. Thus, the linear experiment supports the
same qualitative conclusion as the nonlinear case: temporal information
improves inference for dynamically active edges but does not eliminate
the identifiability limitations associated with weak excitation.

\begin{figure}[htb]
	\centering
	\includegraphics[width=1\linewidth]
	{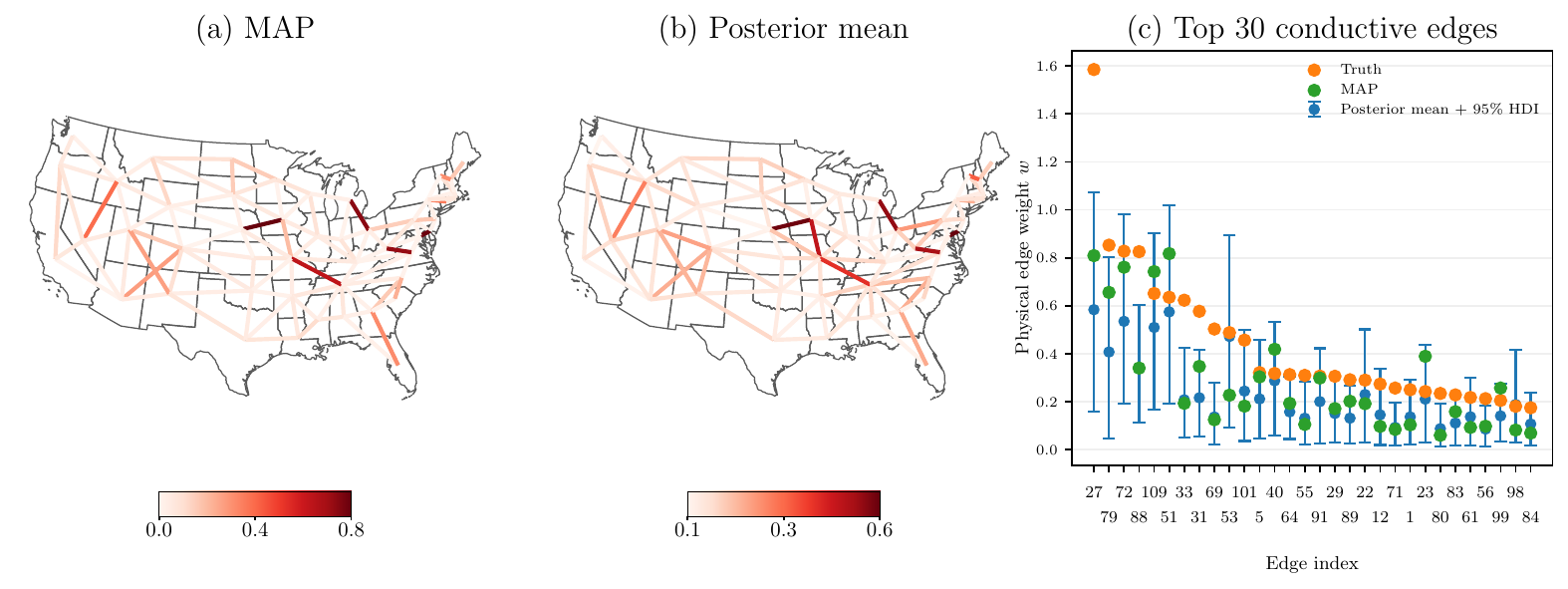}
	\caption{Graph-weight inference for the synthetic non-stationary
		reaction--diffusion problem with linear reaction dynamics:
		(a) MAP estimate, (b) posterior mean, and
		(c) posterior edge-wise uncertainty.}
	\label{fig:sm-heat-linearreaction}
\end{figure}

\subsection{Additional results for the nonstationary graph reaction–diffusion dynamics of COVID-19 in the United States}

We also apply the non-stationary model with the
linearized reaction term to the U.S. state-level COVID-19 data described
in the main text. The model and inference configuration are identical
to those used for the nonlinear reaction model in the main text, with
the reaction term replaced by its linearized counterpart.

The inferred graph weights are shown in
\Cref{fig:sm-heat-covid-linearreaction}. As in the nonlinear model, the
MAP and posterior mean indicate heterogeneous effective inter-state
couplings, with a relatively small number of strongly conductive edges
emerging from the lower-conductivity background. The distinction between
the most conductive edges and the remaining graph is less pronounced
than for the nonlinear model reported in the main text. Nevertheless,
the linear model exhibits qualitatively similar heterogeneity, indicating
that this feature is not specific to the nonlinear reaction term.

\begin{figure}[t!]
	\centering
	\includegraphics[width=1\linewidth]
	{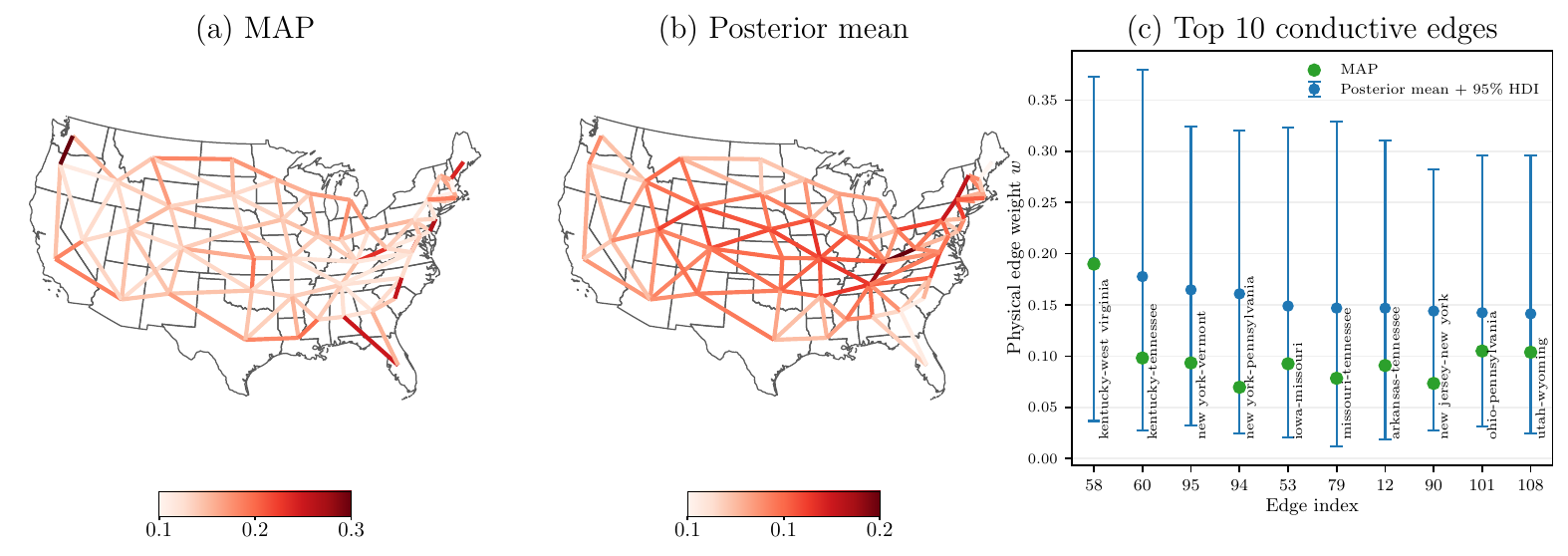}
	\caption{Graph-weight inference from the COVID-19 data using the
		non-stationary linear reaction--diffusion model:
		(a) MAP estimate, (b) posterior mean, and
		(c) posterior edge-wise uncertainty.}
	\label{fig:sm-heat-covid-linearreaction}
\end{figure}

\end{document}